\documentclass[journal]{IEEEtran}
\usepackage{amsfonts}
\usepackage{graphicx}
\usepackage{epstopdf}
\usepackage{amsmath}
\usepackage{amssymb}
\usepackage{algorithmic}
\usepackage{array}
\usepackage{url}
\usepackage{cite}
\usepackage{subfigure}
\usepackage{graphicx}
\usepackage{booktabs}
\usepackage{multirow}
\usepackage{threeparttable}
\usepackage{mathtools}
\usepackage{algorithm}
\usepackage{stfloats}
\usepackage{float}
\usepackage{color}
\usepackage{setspace}

\begin{document}

\title{Coherent FDA Receiver and Joint Range-Space-Time Processing}
\author{ % <-this % stops a space
	Wenkai Jia, Andreas Jakobsson, {\em Senior Member, IEEE}, Wen-Qin Wang, {\em Senior Member, IEEE}
	
	\thanks{This work was supported in part by  National Natural Science Foundation of China 62171092.}
	\thanks{Wenkai Jia and Wen-Qin Wang are School of Information and
		Communication Engineering, University of Electronic Science and
		Technology of China, Chengdu, 611731, P. R. China. (e-mail: wenkai.jia@matstat.lu.se; wqwang@uestc.edu.cn).
		
		Andreas Jakobsson is with the Division of Mathematical Statistics,
		Center for Mathematical Sciences, Lund University, SE-22100 Lund,
		Sweden. (e-mail: andreas.jakobsson@matstat.lu.se).}
}

%\markboth{}%

\maketitle

\begin{abstract}
When a target is masked by mainlobe clutter with the same Doppler frequency, it is difficult for conventional airborne radars to determine whether a target is present in a given observation using regular space-time adaptive processing techniques.
Different from phased-array and multiple-input multiple-output (MIMO) arrays, frequency diverse arrays (FDAs) employ frequency offsets across the array elements, delivering additional range-controllable degrees of freedom, potentially enabling suppression for this kind of clutter.
However, the reception of coherent FDA systems employing small frequency offsets and achieving high transmit gain can be further improved.
To this end, this work proposes an coherent airborne FDA radar receiver that explores the orthogonality of echo signals in the Doppler domain, allowing a joint space-time processing module to be deployed to separate the aliased returns.
The resulting range-space-time adaptive processing allows for a preferable detection performance for coherent airborne FDA radars as compared to current alternative techniques.

\end{abstract}


\section{Introduction}
\IEEEPARstart{M}{oving} target indication (MTI) techniques are used for ground-based radar to separate moving targets and stationary ground clutter by performing band-stop filtering in the time domain or in the Doppler domain.
Weighting each element, spatial filtering enables the array antenna to enhance the desired signal while rejecting unwanted interfering signals from a given direction of arrival (DOA).
However, the ground clutter seen by an airborne radar spreads within the Doppler region due to the platform motion, making it more difficult to decide whether a target is present in a given observation as potential targets may be masked by mainlobe clutter with the same angle as the target and by sidelobe clutter from the same Doppler frequency \cite{skolnik1980introduction}.
One may employ space-time adaptive processing (STAP) using spatial and temporal degrees of freedoms in order to adjust the two-dimensional space-time receiving filter weights in an attempt at maximizing the filter’s output SINR, thereby improving the radar detection capacity \cite{klemm2002principles}. Regrettably, when the target is in the mainlobe of the antenna, although the clutter spectrum is basically a narrow ridge, slow targets will fall into the resulting stop-band, thereby causing performance degradation.

The use of frequency diverse arrays (FDAs), first proposed by Antonic \emph{et al.} \cite{1631800,8321488}, using small frequency offsets (FOs) across the elements, has proven to be advantageous in mainlobe interference suppression \cite{9161264,gui2021fda}, joint estimation of range and angle \cite{6737322,7084678,8049352}, and target tracking \cite{7422108,gui2021cognitive}.
This gain result from the possibility to also exploit the available and controllable degrees of freedom in the range dimension.
From a transmission perspective, the key aspect of FDA is that a range (or delay) increment along a certain direction translates into a relative phase progression among identical sources, making it possible to achieve a focused beampattern in the desired range-angle plane by designing the FOs appropriately \cite{6509929,2016Frequency,2017Joint,WANG201914}.
However, the processing of the FDA echoes is involved due to the need to separate the resulting aliased multi-carrier signals.
This difficulty may be avoided in the case when the FO is larger than the bandwidth of the baseband waveform \cite{8448961,8954891,lan2021single}.
Since the spectra do not overlap in this case, a low-pass filter is then able to separate the components without difficulty, although this gain comes at the price of an increased bandwidth.
For a co-located MIMO, which transmits mutually orthogonal waveforms, the reception procedure used by an FDA-MIMO cause a loss in the resulting transmission gain.
Various approaches have been introduced to handle the received signals, such as the  
multichannel matched filtering structure presented in \cite{8074796}, where each receive antenna is implemented by a group of carriers with matched filters. These may be applied to FDA systems using arbitrary FOs, including linear or nonlinear ones.
However, for small FO, the necessary requirement of waveform orthogonality for all Doppler and delay pairs makes such a solution infeasible in practice.

In this paper, we focus on the application of range-space-time processing for airborne coherent FDA radars, introducing a receiver that exploits the orthogonality in the Doppler domain.
As even small FOs will destroy orthogonality and the coherence of the waveforms, the fast-time receiving structure will not be able to deal with the aliased returns.
Different from the multi-channel filtering or the low-pass filtering receiver, where the returns are only processed in the fast-time, the here proposed receiver jointly filters the returns in the space-time domain.
We initially consider a pulsed-FDA transmit signal with a coded phase for each element, although it is worth noting that the here introduced phase code is different from the space-time coding technique in \cite{9573483}, which aims at shifting the frequency band of each pulse to synthesize a full bandwidth, thereby improving the range resolution.
The here proposed phase encoding scheme instead imparts an additional Doppler shift to the pulsed signal transmitted by each antenna.
Then, in the receiver, after processing the received signal using multi-carrier mixing and applying the matched filtering in the fast-time domain, the subsequent steps include the Doppler demodulation and a low-pass filtering in the slow-time, i.e., the inter-pulse time domain.
The resulting joint space-time  processing is capable of separating the aliased returns without any additional constraints on the FO or on the transmit waveforms, yielding a coherent FDA range-space-time signal model.
Numerical simulations demonstrate the effectiveness of proposed receiver for coherent FDA and an achievable high-gain detection of slow moving targets that would be obscured by the clutter ridge of a conventional airborne STAP radar.

The remainder of the paper is organized as follows.
In Section \uppercase\expandafter{\romannumeral2}, we briefly review conventional receiving schemes and their suitability for FDA-MIMO and for coherent FDA array systems.
In Section \uppercase\expandafter{\romannumeral3}, we introduce the proposed FDA receiver based on joint space-time processing to efficiently receive FDA returns.
In Section \uppercase\expandafter{\romannumeral4}, considering an airborne FDA radar, the resulting interference and noise models are analyzed, and a weight-based FDA range-space-time processing scheme is developed.
Simulation results are given in Section \uppercase\expandafter{\romannumeral5}. 
Finally, in Section \uppercase\expandafter{\romannumeral6}, we present our conclusions.

\section{Background}
Consider a uniformly spaced linear FDA radar with $N_T$ transmit antennas measuring the backscattering resulting from transmitters using an uniform FO with central carrier frequency $f_c$, each of which emits the same narrowband waveform, $u\left( t \right)$, having unit energy, i.e., $\int_{{{T}_{p}}}{{{\left| u\left( t \right) \right|}^{2}}}\operatorname{d}t=1$, where $T_p$ denotes the pulse duration.
The transmitted signal, $s\left( t \right)$, may then be expressed as
\begin{equation}
s\left( t \right)=\sum\limits_{m=1}^{{{N}_{T}}}{{{w}_{m}}u\left( t \right){{e}^{j2\pi \left( {{f}_{c}}+\left( m-1 \right)\Delta f \right)t}}}
\end{equation}
where $w_m$ and $\Delta f$ denote the transmit weight of the $m$-th antenna and the FO, respectively.
Assuming a far-field point target located at the angle-range pair $\left( {{\theta }_{t}},{{r}_{t}} \right)$, the reflected signal, $\bar{s}\left( t \right)$, may be expressed as
\begin{equation}
\begin{aligned}
\bar{s}\left( t \right)& =s\left( t-\left( \frac{{{r}_{t}}}{c}-\left( m-1 \right)\frac{{{d}_{t}}\sin \theta }{c} \right) \right) \\ 
& =\sum\limits_{m=1}^{{{N}_{T}}}{\left\{ \begin{aligned}
	& {{w}_{m}}u\left( t-\left( \frac{{{r}_{t}}}{c}-\left( m-1 \right)\frac{{{d}_{t}}\sin \theta }{c} \right) \right) \\ 
	& \times {{e}^{j2\pi \left( {{f}_{c}}+\left( m-1 \right)\Delta f \right)\left( t-\frac{{{r}_{t}}}{c}+\left( m-1 \right)\frac{{{d}_{t}}\sin \theta }{c} \right)}} \\ 
	\end{aligned} \right\}} \\ 
& \approx {{e}^{j2\pi {{f}_{c}}\left( t-\frac{{{r}_{t}}}{c} \right)}}u\left( t-\frac{{{r}_{t}}}{c} \right) \\ 
& \times \sum\limits_{m=1}^{{{N}_{T}}}{{{w}_{m}}{{e}^{j2\pi \left( m-1 \right)\Delta f\left( t-\frac{{{r}_{t}}}{c} \right)}}{{e}^{j2\pi \left( m-1 \right)\left( {{f}_{c}}+\Delta f \right)\frac{{{d}_{t}}\sin \theta }{c}}}} \\ 
\end{aligned}
\end{equation}
where $d_t$ denotes the spacing between the transmit antennas.
For an FDA receiver with $N_R$ equally spaced receive antennas, the resulting backscattered signal arriving at its $n$-th antenna may then be approximated as
\begin{equation}
\begin{aligned}
{{y}_{n}}\left( t \right)&={{\delta }_{t}}\bar{s}\left( t-\left( \frac{{{r}_{t}}}{c}-\left( n-1 \right)\frac{{{d}_{r}}\sin \theta }{c} \right) \right) \\ 
& \approx {{\delta }_{t}}{{e}^{j2\pi {{f}_{c}}\left( t-\frac{2{{r}_{t}}}{c} \right)}}{{e}^{j2\pi \left( n-1 \right){{f}_{c}}\frac{{{d}_{r}}\sin \theta }{c}}}u\left( t-\frac{2{{r}_{t}}}{c} \right) \\ 
& \times \sum\limits_{m=1}^{{{N}_{T}}}{{{w}_{m}}{{e}^{j2\pi \left( m-1 \right)\Delta f\left( t-\frac{2{{r}_{t}}}{c} \right)}}{{e}^{j2\pi \left( m-1 \right)\left( {{f}_{c}}+\Delta f \right)\frac{{{d}_{t}}\sin \theta }{c}}}} \\ 
\end{aligned}
\end{equation}
where ${{\delta }_{t}}$ and $d_r$ denote the target's complex-valued reflection coefficient and the spacing between the receive antennas, respectively.
It should be noted that ${{\delta }_{t}}$ may be decorrelated in frequency as a result of the FDA's multi-carrier transmission mechanism, although a sufficient condition to avoid the frequency decorrelation for an FDA radar with linear FO is \cite{9266663}
\begin{equation}
\Delta f\le \frac{c}{4\left( {{N}_{T}}-1 \right)\Delta \varsigma }
\end{equation}
where $\Delta \varsigma$ denotes the target length along the radar's boresight.

\begin{figure}
	\centering
	\includegraphics[width=0.45\textwidth]{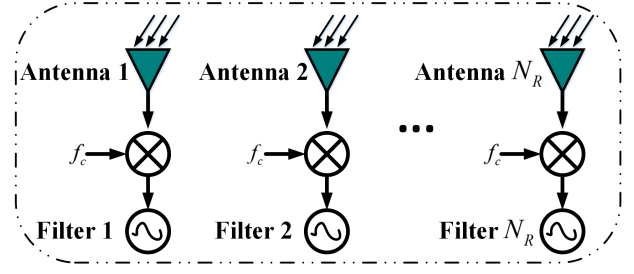}
	\caption{The receiver for FDA-MIMO system.}
\end{figure}
Fig. 1 shows an FDA-MIMO system, i.e., a receiver structure fully utilizing the orthogonality of the signal in the frequency domain in the case the FO is larger than the baseband bandwidth. 
For each receive antenna, the returned FDA signals are demodulated using the central carrier frequency, $f_c$, followed by a suitable band-pass filter. 
As a result, the output of the $n$-th receive antenna can be expressed as
\begin{equation}
\begin{aligned}
{{{\bar{y}}}_{n}}\left( t \right)&=\left( {{y}_{n}}\left( t \right)\cdot {{e}^{-j2\pi {{f}_{c}}t}} \right)*h_{n}^{c}\left( t \right) \\ 
& ={{\delta }_{t}}{{e}^{-j2\pi {{f}_{c}}\frac{2{{r}_{t}}}{c}}}{{e}^{j2\pi \left( n-1 \right){{f}_{c}}\frac{{{d}_{r}}\sin \theta }{c}}} \\ 
& \times \left[ {{\kappa }_{n}}\left( t \right)*h_{n}^{c}\left( t \right)+\underbrace{\left( \sum\limits_{\begin{smallmatrix} 
		m=1 \\ 
		m\ne n 
		\end{smallmatrix}}^{{{N}_{T}}}{{{\kappa }_{m}}\left( t \right)} \right)*h_{n}^{c}\left( t \right)}_\text{aliased \kern 2pt term} \right] \\ 
& ={{\delta }_{t}}{{e}^{-j2\pi {{f}_{c}}\frac{2{{r}_{t}}}{c}}}{{e}^{j2\pi \left( n-1 \right){{f}_{c}}\frac{{{d}_{r}}\sin \theta }{c}}}{{\kappa }_{n}}\left( t \right) \\ 
\end{aligned}
\end{equation}
where
\begin{equation}
\begin{aligned}
{{\kappa }_{n}}\left( t \right)&={{w}_{n}}{{e}^{j2\pi \left( n-1 \right)\left( {{f}_{c}}+\Delta f \right)\frac{{{d}_{t}}\sin \theta }{c}}} \\ 
& \kern 6pt \times u\left( t-\frac{2{{r}_{t}}}{c} \right){{e}^{j2\pi \left( n-1 \right)\Delta f\left( t-\frac{2{{r}_{t}}}{c} \right)}}, \\ 
\end{aligned}
\end{equation}
with $*$ and ${{\left( \cdot  \right)}^{c}}$ denoting the convolution and the conjugate operators, respectively, and where ${{h}_{n}}\left( t \right)$ is the used band-pass filter, having passband ${{f}_{pass}}\in \left[ \left( n-1 \right)\Delta f,n\Delta f \right]$.
Since the FO is larger than the baseband bandwidth, the occupied spectra of the aliased term do not overlap with that of ${{\kappa }_{n}}\left( t \right)*h_{n}^{c}\left( t \right)$ in $(5)$, implying that the filter is capable of extracting the desired signal from the aliased ones.
It is worth noting that the receiver uses the same number of receive antennas as transmit antennas, indicating that the virtual array may not be formed.

For the case where the FO is smaller than the baseband bandwidth, \cite{8074796} proposed a multi-carrier mixing and matched filtering receiver structure.
First, the echo signal to the $n$-th receive antenna, ${{y}_{n}}\left( t \right)$, is demodulated by multi-channel mixers with local carrier frequencies $\left\{ {{f}_{c}}+\left( m-1 \right)\Delta f \right\}_{m=1}^{{{N}_{T}}}$, which is then matched filtered with the baseband waveform, $u\left( t \right)$, to obtain the pulse compression gain. 
The filter output of $m'$-th channel of the $n$-th receive antenna may as a result be expressed as  
\begin{equation}
\begin{aligned}
{{{\tilde{y}}}_{n,{m}'}}\left( t \right) & =\left( {{y}_{n}}\left( t \right)\cdot {{e}^{-j2\pi \left( {{f}_{c}}+\left( {m}'-1 \right)\Delta f \right)t}} \right)*{{u}^{c}}\left( -t \right) \\ 
& ={{\delta }_{t}}{{o}_{n,{m}'}}\int_{{{T}_{p}}}{u\left( \tau -\frac{2{{r}_{t}}}{c} \right){{u}^{c}}\left( \tau -t \right)\text{d}\tau }+{{\delta }_{t}} \\ 
& \kern -43pt \times \underbrace{\sum\limits_{\begin{smallmatrix} 
		m=1 \\ 
		m\ne {m}' 
		\end{smallmatrix}}^{{{N}_{T}}}{{{o}_{n,m}}\int_{{{T}_{p}}}{{{e}^{j2\pi \left( m-{m}' \right)\Delta f\tau }}u\left( \tau -\frac{2{{r}_{t}}}{c} \right){{u}^{c}}\left( \tau -t \right)\text{d}\tau }}}_\text{aliased \kern 2pt term} \\ 
\end{aligned}
\end{equation}
where 
\begin{equation}
\begin{aligned}
{{o}_{n,{m}'}}& ={{w}_{{{m}'}}}{{e}^{-j2\pi {{f}_{c}}\frac{2{{r}_{t}}}{c}}}{{e}^{j2\pi \left( n-1 \right){{f}_{c}}\frac{{{d}_{r}}\sin \theta }{c}}} \\ 
& \kern 6pt \times {{e}^{j2\pi \left( {m}'-1 \right)\left( \left( {{f}_{c}}+\Delta f \right)\frac{{{d}_{t}}\sin \theta }{c}-\Delta f\frac{2{{r}_{t}}}{c} \right)}} \\ 
\end{aligned}
\end{equation}
It is worth noting that if
\begin{equation}
\begin{aligned}
& \int_{{{T}_{p}}}{{{e}^{j2\pi \left( m-{m}' \right)\Delta f\tau }}u\left( \tau -\frac{2{{r}_{t}}}{c} \right){{u}^{c}}\left( \tau -t \right)\text{d}\tau }=0, \\ 
& \kern 165pt \forall t,m\ne {m}' \\ 
\end{aligned}
\end{equation}
the aliased term in $(7)$ will be filtered, yielding the desired output
\begin{equation}
{{\tilde{y}}_{n,{m}'}}\left( t \right)={{\delta }_{t}}{{o}_{n,{m}'}}\int_{{{T}_{p}}}{u\left( \tau -\frac{2{{r}_{t}}}{c} \right){{u}^{c}}\left( \tau -t \right)\text{d}\tau }
\end{equation}
However, it may be seen that the orthogonality condition in $(9)$ is so strict that it is impossible to find a suitable waveform $u\left( t \right)$ and FO $\Delta f$, although it is noted in \cite{8074796} that for extremely small FOs, the condition can be approximately satisfied, but then with an unavoidable performance loss.

\section{Proposed receiver for coherent FDA}
The assumed system is a pulsed-Doppler FDA radar on an airborne platform moving with constant velocity $v_a$, as illustrated in Fig. 2.
The $N_T$ transmit antennas and the $N_R$ receive antennas are uniformly arranged along the movement direction of the platform, forming a side-looking airborne FDA radar.
In the figure, $\psi $ denotes the conic angle, which satisfies
\begin{equation}
\cos \left( \psi  \right)=\cos \left( \theta  \right)\cos \left( \phi  \right)
\end{equation}
with $\theta $ and $\phi $ representing the azimuth angle and the depression angle, respectively.

\begin{figure}
	\centering
	\includegraphics[width=0.45\textwidth]{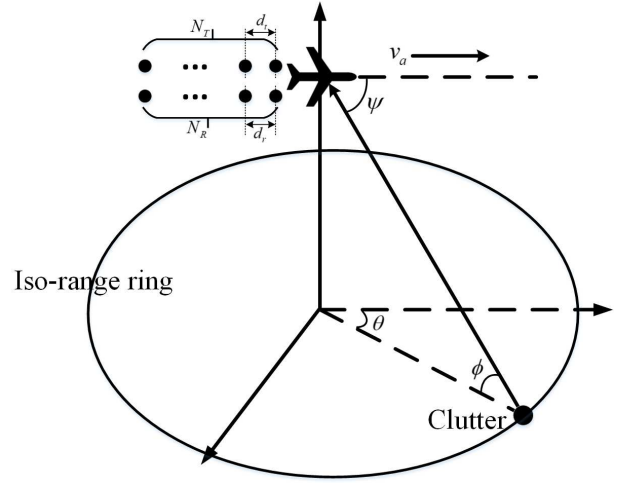}
	\caption{Geometric model of airborne FDA radar antenna.}
\end{figure}

\subsection{Transmit Signal Model }
The proposed receiver is based on the orthogonality of the received signal in the Doppler domain in order to facilitate the channel separation during pulse compression.
Firstly, for each transmit antenna, we introduce the designed phase codes ${{\varphi }_{m}}$, for $m=1,2,...,{{N}_{T}}$.
Thus, the $l$-th pulse transmitted by the $N_T$ antennas can be expressed as
\begin{equation}
s\left( {{t}_{l}},\tau  \right)=\sum\limits_{m=1}^{{{N}_{T}}}{{{w}_{m}}\operatorname{rect}\left( \frac{\tau }{{{T}_{p}}} \right)u\left( t \right){{e}^{j2\pi \left( \left( {{f}_{c}}+\left( m-1 \right)\Delta f \right)t+{{\varphi }_{m}}{{t}_{l}} \right)}}}
\end{equation}
for $l=1,2,...,L$, with $L$ denoting the number of pulses, where
\begin{equation}
\operatorname{rect}\left( \frac{\tau }{{{T}_{p}}} \right)=\left\{ \begin{matrix}
1 & \left| \tau  \right|\le \frac{{{T}_{p}}}{2}  \\
	0 & \left| \tau  \right|>\frac{{{T}_{p}}}{2}  \\
\end{matrix} \right.,
\end{equation}
and with $\tau =t -{{t}_{l}}$ and ${{t}_{l}}=l{{T}_{r}}$
denoting the fast and slow time, respectively, with $T_r$ being the pulse repetition interval (PRI). 
It is worth noting that the resulting FDA signal is composed of $N_T$ signals with different carrier frequencies $\left\{ {{f}_{c}}+\left( m-1 \right)\Delta f \right\}_{m=1}^{{{N}_{T}}}$, each signal consisting of $L$ pulses with induced Doppler ${{\varphi }_{m}}$.
\subsection{Echo Signal Model }
The target is defined as a point scatterer moving with velocity $v_t$ located at the range-angle pair $\left( {{r}_{t}},{{\theta }_{t}},{{\phi }_{t}} \right)$.
The target delay ${{\xi }_{m,n,l}}$ consists of four components, such that
\begin{equation}
\begin{aligned}
{{\xi }_{m,n,l}}&=\xi \left( {{r}_{t}} \right)+\left( m-1 \right){{\xi }_{T}}\left( {{\psi }_{t}} \right) \\ 
& +\left( n-1 \right){{\xi }_{R}}\left( {{\psi }_{t}} \right)+{{\xi }_{D}}\left( \tau  \right)+{{\xi }_{D}}\left( {{t}_{l}} \right) \\ 
\end{aligned}
\end{equation}
where
\begin{equation}
\xi \left( {{r}_{t}} \right)=\frac{2{{r}_{t}}}{c}
\end{equation}
denotes the round trip delay, 
\begin{subequations}
	\begin{equation}
{{\xi }_{T}}\left( {{\psi }_{t}} \right)=-\frac{{{d}_{t}}\cos \left( {{\theta }_{t}} \right)\cos \left( {{\phi }_{t}} \right)}{c}=-\frac{{{d}_{t}}\cos {{\psi }_{t}}}{c}
	\end{equation}
	\begin{equation}
{{\xi }_{R}}\left( {{\psi }_{t}} \right)=-\frac{{{d}_{r}}\cos \left( {{\theta }_{t}} \right)\cos \left( {{\phi }_{t}} \right)}{c}=-\frac{{{d}_{r}}\cos {{\psi }_{t}}}{c}
	\end{equation}
\end{subequations}
represent the relative delay between the transmit antennas and the relative delay between the receive antennas, respectively, and
\begin{subequations}
	\begin{equation}
{{\xi }_{D}}\left( \tau  \right)=\frac{2{{v}_{t}}\tau }{c}\times {{f}_{c}}\times \frac{1}{{{f}_{c}}}=\frac{{{f}_{t,d}}}{{{f}_{c}}}\tau 
	\end{equation}
	\begin{equation}
\kern -79pt {{\xi }_{D}}\left( {{t}_{l}} \right)=\frac{{{f}_{t,d}}}{{{f}_{c}}}{{t}_{l}}
	\end{equation}
\end{subequations}
are the delays of the target within and between pulses, respectively, where ${{f}_{t,d}}=\frac{2{{v}_{t}}}{\lambda }$ denotes the target Doppler with $\lambda =\frac{c}{{{f}_{c}}}$ being the wavelength.

In order to obtain the analytical echo signal model, three assumptions are made, namely
\begin{itemize}
	\item[1)] The radar cross-section (RCS) of the targets within the same coherent processing interval (CPI) do not fluctuate significantly, i.e., the target obeys a Swerling \uppercase\expandafter{\romannumeral3} model \cite{richards2014fundamentals}.
	\item[2)] The envelope delays due to the antenna spacing and the target motion are negligible\footnote{High-speed maneuvering targets will cause range migration (RM) and Doppler frequency migration (DFM) resulting in that a target will be more difficult to detect. We note that long-time coherent integration by compensating range walk and phase modulations over different sampling pulses is an effective method to improve radar target detection performance \cite{xu2011radon,9347707}}.
	\item[3)] The pulse duration is small enough so that the intrapulse Doppler shift is negligible.
\end{itemize}
As a result, the $l$-th pulse transmitted by the $m$-th transmit antenna reflected to the $n$-th receive antenna has the form
\begin{equation}
\begin{aligned}
{{y}_{m,n}}\left( {{t}_{l}},\tau  \right)&={{\delta }_{t}}{{w}_{m}}\operatorname{rect}\left( \frac{\tau -{{\xi }_{m,n,l}}}{{{T}_{p}}} \right)u\left( t-{{\xi }_{m,n,l}} \right) \\ 
& \kern 10pt \times {{e}^{j2\pi \left( \left( {{f}_{c}}+\left( m-1 \right)\Delta f \right)\left( t-{{\xi }_{m,n,l}} \right)+{{\varphi }_{m}}{{t}_{l}} \right)}} \\ 
& \approx {{\delta }_{t}}{{a}_{m,n}}\left( {{\psi }_{t}} \right){{w}_{m}}\operatorname{rect}\left( \frac{\tau -\xi \left( r \right)}{{{T}_{p}}} \right)u\left( t-\xi \left( {{r}_{t}} \right) \right) \\ 
& \kern 10pt \times {{e}_{m}}\left( t-\xi \left( {{r}_{t}} \right) \right){{\omega }_{D,m}}\left( {{t}_{l}},{{\varphi }_{m}} \right) \\ 
\end{aligned}
\end{equation}
where
\begin{subequations}
\begin{equation}
\kern 12pt {{a}_{m,n}}\left( {{\psi }_{t}} \right)={{e}^{j2\pi \left( {{f}_{c}}+\left( m-1 \right)\Delta f \right)\left[ \left( m-1 \right){{\xi }_{T}}\left( {{\psi }_{t}} \right)+\left( n-1 \right){{\xi }_{R}}\left( {{\psi }_{t}} \right) \right]}}
\end{equation}
\begin{equation}
\kern -75pt {{e}_{m}}\left( t \right)={{e}^{j2\pi \left( {{f}_{c}}+\left( m-1 \right)\Delta f \right)t}}
\end{equation}
\begin{equation}
\kern -62pt \begin{aligned}
{{\omega }_{D,m}}\left( {{t}_{l}},{{\varphi }_{m}} \right)&={{e}^{j2\pi \left[ \left( {{f}_{c}}+\left( m-1 \right)\Delta f \right){{\xi }_{D}}\left( {{t}_{l}} \right)+{{\varphi }_{m}}{{t}_{l}} \right]}} \\ 
& ={{e}^{j2\pi \left( \frac{\left( {{f}_{c}}+\left( m-1 \right)\Delta f \right)}{{{f}_{c}}}{{f}_{t,d}}+{{\varphi }_{m}} \right){{t}_{l}}}}. \\ 
\end{aligned}
\end{equation}
\end{subequations}
This implies that the sum of all $N_T$ signals reflected to the $n$-th receive antenna may be expressed as 
\begin{equation}
\begin{aligned}
{{y}_{n}}\left( {{t}_{l}},\tau  \right)&=\sum\limits_{m=1}^{{{N}_{T}}}{{{y}_{m,n}}\left( {{t}_{l}},\tau  \right)} \\ 
& ={{\delta }_{t}}\mathbf{a}_{n}^{T}\left( {{\psi }_{t}} \right)\mathbf{\Omega }\left( {{t}_{l}},\varphi  \right)\mathbf{U}\left( \tau ,\xi \left( {{r}_{t}} \right);\mathbf{w} \right)\mathbf{e}\left( t-\xi \left( {{r}_{t}} \right) \right) \\ 
\end{aligned}
\end{equation}
where
\begin{subequations}
	\begin{equation}
	\kern -96pt \mathbf{\Omega }\left( {{t}_{l}},\varphi  \right)=\operatorname{diag}\left\{ {{\boldsymbol{\omega }}_{D}}\left( {{t}_{l}},\varphi  \right) \right\}
	\end{equation}
	\begin{equation}
	\begin{aligned}
	& {{\boldsymbol {\omega} }_{D}}\left( {{t}_{l}},\varphi  \right)= \\ 
	& {{\left[ {{\omega }_{D,1}}\left( {{t}_{l}},{{\varphi }_{1}} \right),{{\omega }_{D,2}}\left( {{t}_{l}},{{\varphi }_{2}} \right),...,{{\omega }_{D,{{N}_{T}}}}\left( {{t}_{l}},{{\varphi }_{{{N}_{T}}}} \right) \right]}^{T}}, \\ 
	\end{aligned}
	\end{equation}
\end{subequations}
with $\operatorname{diag}\left\{ {{\boldsymbol{\omega }}_{D}}\left( {{t}_{l}},\varphi  \right) \right\}$ denoting the diagonal matrix with entries formed by ${{\boldsymbol{\omega }}_{D}}\left( {{t}_{l}},\varphi  \right)$, and where
\begin{subequations}
	\begin{equation}
\kern 22pt {{\mathbf{a}}_{n}}\left( {{\psi }_{t}} \right)={{\left[ {{a}_{1,n}}\left( {{\psi }_{t}} \right),{{a}_{2,n}}\left( {{\psi }_{t}} \right),...,{{a}_{{{N}_{T}},n}}\left( {{\psi }_{t}} \right) \right]}^{T}}
	\end{equation}
		\begin{equation}
	\kern -4pt \begin{aligned}
	& \mathbf{e}\left( t-\xi \left( {{r}_{t}} \right) \right)= \\ 
	& {{\left[ {{e}_{1}}\left( t-\xi \left( {{r}_{t}} \right) \right),{{e}_{2}}\left( t-\xi \left( {{r}_{t}} \right) \right),...,{{e}_{{{N}_{T}}}}\left( t-\xi \left( {{r}_{t}} \right) \right) \right]}^{T}}, \\ 
	\end{aligned}
	\end{equation}
\end{subequations}
	\begin{equation}
\mathbf{U}\left( \tau ,\xi \left( {{r}_{t}} \right);\mathbf{w} \right)=\operatorname{rect}\left( \frac{\tau -\xi \left( {{r}_{t}} \right)}{{{T}_{p}}} \right)u\left( t-\xi \left( {{r}_{t}} \right) \right)\operatorname{diag}\left\{ \mathbf{w} \right\}
\end{equation}
with
\begin{equation}
\mathbf{w}={{\left[ {{w}_{1}},{{w}_{2}},...,{{w}_{{{N}_{T}}}} \right]}^{T}}
\end{equation}
denoting the transmit weight vector.

\subsection{Multi-carrier Mixing}
To separate the desired signal from the aliased ones, the proposed receiver initially forms a multi-carrier mixing in the fast-time domain.
Similar to the receiver in \cite{8074796}, the used mixing frequencies are $\left\{ {{f}_{c}}+\left( m-1 \right)\Delta f \right\}_{m=1}^{{{N}_{T}}}$. 
Thus, the mixed output of the $m'$-th channel may be expressed as
\begin{equation}
\begin{aligned}
{{r}_{{m}',n}}\left( {{t}_{l}},\hat{\tau } \right)&={{y}_{n}}\left( {{t}_{l}},\tau  \right){{e}^{-j2\pi \left( {{f}_{c}}+\left( {m}'-1 \right)\Delta f \right)t}} \\ 
& ={{\delta }_{t}}\mathbf{a}_{n}^{T}\left( \xi \left( {{r}_{t}} \right),{{\psi }_{t}} \right){{\mathbf{\Omega }}^{{{m}'}}}\left( {{t}_{l}},\varphi  \right){{\mathbf{u}}^{{{m}'}}}\left( \tau ,\xi \left( {{r}_{t}} \right);\mathbf{w} \right) \\ 
\end{aligned}
\end{equation}
for ${m}'=1,2,...,{{N}_{T}}$, where
\begin{subequations}
	\begin{equation}
\kern -83pt {{\mathbf{a}}_{n}}\left( \xi \left( {{r}_{t}} \right),{{\psi }_{t}} \right)={{\mathbf{a}}_{n}}\left( {{\psi }_{t}} \right)\odot \mathbf{e}\left( \xi \left( {{r}_{t}} \right) \right)
	\end{equation}
	\begin{equation}
\begin{aligned}
& \kern 11pt \mathbf{e}\left( \xi \left( {{r}_{t}} \right) \right)= \\ 
& {{\left[ {{e}_{1}}\left( -\xi \left( {{r}_{t}} \right) \right),{{e}_{2}}\left( -\xi \left( {{r}_{t}} \right) \right),...,{{e}_{{{N}_{T}}}}\left( -\xi \left( {{r}_{t}} \right) \right) \right]}^{T}}, \\ 
\end{aligned}
	\end{equation}
\end{subequations}
with $\odot$ denoting the Hadamard matrix product,
\begin{subequations}
	\begin{equation}
	\kern -114pt {{\mathbf{\Omega }}^{{{m}'}}}\left( {{t}_{l}},\varphi  \right)=\operatorname{diag}\left\{ {{\boldsymbol{\omega }}^{{{m}'}}}\left( {{t}_{l}},\varphi  \right) \right\}
	\end{equation}
	\begin{equation}
	\kern -106pt \begin{aligned}
	{{\boldsymbol{\omega }}^{{{m}'}}}\left( {{t}_{l}},\varphi  \right)&=\boldsymbol{\Omega }\left( {{t}_{l}},\varphi  \right){{\mathbf{e}}^{{{m}'}}}\left( {{t}_{l}} \right) \\ 
	& ={{\boldsymbol{\omega }}_{D}}\left( {{t}_{l}},\varphi  \right)\odot {{\mathbf{e}}^{{{m}'}}}\left( {{t}_{l}} \right) \\ 
	\end{aligned}
	\end{equation}
	\begin{equation}
	\kern 12pt \begin{aligned}
	& {{\mathbf{e}}^{{{m}'}}}\left( {{t}_{l}} \right)= \\ 
	& {{\left[ {{e}^{j2\pi \left( 1-{m}' \right)\Delta f{{t}_{l}}}},{{e}^{j2\pi \left( 2-{m}' \right)\Delta f{{t}_{l}}}},...,{{e}^{j2\pi \left( {{N}_{T}}-{m}' \right)\Delta f{{t}_{l}}}} \right]}^{T}}, \\ 
	\end{aligned}
	\end{equation}
\end{subequations}
and
\begin{subequations}
	\begin{equation}
	\kern 28pt \begin{aligned}
	&{{\mathbf{u}}^{{{m}'}}}\left( \tau ;\xi \left( {{r}_{t}} \right);\mathbf{w} \right)\\
	&=\mathbf{U}\left( \tau ;\xi \left( {{r}_{t}} \right);\mathbf{w} \right){{\mathbf{e}}^{{{m}'}}}\left( \tau  \right) \\ 
	& =\operatorname{rect}\left( \frac{\tau -\xi \left( {{r}_{t}} \right)}{{{T}_{p}}} \right)u\left( \tau -\xi \left( {{r}_{t}} \right) \right)\mathbf{w}\odot {{\mathbf{e}}^{{{m}'}}}\left( \tau  \right) \\ 
	\end{aligned}
	\end{equation}
	\begin{equation}
	\begin{aligned}
	& {{\mathbf{e}}^{{{m}'}}}\left( \tau  \right) =\\ 
	& {{\left[ {{e}^{j2\pi \left( 1-{m}' \right)\Delta f\tau }},{{e}^{j2\pi \left( 2-{m}' \right)\Delta f\tau }},...,{{e}^{j2\pi \left( {{N}_{T}}-{m}' \right)\Delta f\tau }} \right]}^{T}} \\ 
	\end{aligned}
	\end{equation}
\end{subequations}
Here, the symbol $\hat{\tau }$ denotes the processing in the fast-time domain.

\subsection{Matched Filtering}
To obtain the pulse compression gain, one proceeds to form a matched filtering in the fast-time domain, with
the filter output of the $m'$-th channel being the signal
\begin{equation}
\begin{aligned}
{{{\bar{r}}}_{{m}',n}}\left( {{t}_{l}},\hat{\tau } \right)&={{r}_{{m}',n}}\left( {{t}_{l}},\hat{\tau } \right)*h\left( \tau  \right) \\ 
& ={{\delta }_{t}}\mathbf{a}_{n}^{T}\left( \xi \left( {{r}_{t}} \right),{{\psi }_{t}} \right){{\mathbf{\Omega }}^{{{m}'}}}\left( {{t}_{l}},\varphi  \right){{\mathbf{g}}^{{{m}'}}}\left( \tau;\mathbf{w}  \right) \\ 
\end{aligned}
\end{equation}
where
\begin{equation}
h\left( \tau  \right)={{u}^{c}}\left(\tau  \right)
\end{equation}
denotes the matched waveform, and
\begin{equation}
{{\mathbf{g}}^{{{m}'}}}\left( \tau ;\mathbf{w}  \right)=\left[ g_{1}^{{{m}'}}\left( \tau;w_1  \right),g_{2}^{{{m}'}}\left( \tau ;w_2  \right),...,g_{{{N}_{T}}}^{{{m}'}}\left( \tau ;w_{N_T}  \right) \right]
\end{equation}
where
\begin{equation}
\begin{aligned}
& g_{m}^{{{m}'}}\left( \tau;w_m  \right)\\
&={{w}_{m}}\left[ u\left( \tau -\xi \left( {{r}_{t}} \right) \right){{e}^{j2\pi \left( m-{m}' \right)\Delta f\tau }} \right]*h\left( \tau  \right) \\ 
& ={{w}_{m}}\int_{{{T}_{P}}}{u\left( \beta -\xi \left( {{r}_{t}} \right) \right){{e}^{j2\pi \left( m-{m}' \right)\Delta f\beta }}{{u}^{c}}\left( \beta-\tau  \right)\operatorname{d}\beta } \\ 
& ={{w}_{m}}{{e}^{j2\pi \left( m-{m}' \right)\Delta f\xi \left( {{r}_{t}} \right)}} \\ 
& \kern 9pt \times \int_{{{T}_{P}}}{u\left( \beta  \right){{e}^{j2\pi \left( m-{m}' \right)\Delta f\beta }}{{u}^{c}}\left( \beta +\xi \left( {{r}_{t}} \right)-\tau  \right)\operatorname{d}\beta } \\ 
& ={{w}_{m}}\mathsf{\mathbb{G}}\left( \tau -\xi \left( {{r}_{t}} \right),\left( m-{m}' \right)\Delta f \right) \\ 
\end{aligned}
\end{equation}
with
\begin{equation}
\mathsf{\mathbb{G}}\left( {\tau }',{f}' \right)=\int_{{{T}_{P}}}{u\left( \tau  \right){{e}^{j2\pi \tau {f}'}}{{u}^{c}}\left( \tau -{\tau }' \right)\operatorname{d}\tau }
\end{equation}
being the ambiguity function of $u\left( \tau  \right)$, implying that when $m={m}'$, the maximum occurs at the target's round trip delay, i.e., $\tau =\xi \left( {{r}_{t}} \right)$.

\subsection{Doppler Demodulation}
The third step is Doppler demodulation in the slow-time domain, shifting the $m'$-th channel signal to the baseband in the Doppler domain. 
In order to do so, $(29)$ is expressed as
\begin{equation}
{{\bar{r}}_{{m}',n}}\left( {{t}_{l}},\hat{\tau } \right)={{\delta }_{t}}\mathbf{a}_{n}^{T}\left( \xi \left( {{r}_{t}} \right),{{\psi }_{t}} \right){{\mathbf{G}}^{{{m}'}}}\left( \tau; \mathbf{w} \right){{\mathbf{\omega }}^{{{m}'}}}\left( {{t}_{l}},\varphi  \right)
\end{equation}
where
\begin{subequations}
	\begin{equation}
\kern -15pt {{\mathbf{G}}^{{{m}'}}}\left( \tau ; \mathbf{w} \right)=\operatorname{diag}\left\{ {{\mathbf{g}}^{{{m}'}}}\left( \tau ; \mathbf{w} \right) \right\}
	\end{equation}
	\begin{equation}
\kern 20pt {\boldsymbol{\omega }^{{{m}'}}}\left( {{t}_{l}};\varphi  \right)={\boldsymbol{\omega }_{D,{m}'}}\left( {{t}_{l}},{{\varphi }_{{{m}'}}} \right){{e}^{j2\pi {m}'\Delta f{{t}_{l}}}}
	\end{equation}
\end{subequations}
with $\mathbf{a}_{n}\left( \xi \left( {{r}_{t}} \right),{{\psi }_{t}} \right)$ as given in $(26a)$.
Then, the Doppler demodulation may be performed by using the modulated slow-time signal ${{e}^{-j2\pi {{\varphi }_{{{m}'}}}{{t}_{l}}}}$, yielding the demodulated signal
\begin{equation}
\begin{aligned}
& {{{\tilde{r}}}_{{m}',n}}\left( {{{\hat{t}}}_{l}},\tau  \right)=\sum\limits_{m=1}^{{{N}_{T}}}{\left\{ {{{\bar{r}}}_{{m}',n}}\left( {{{\hat{t}}}_{l}},\tau  \right){{e}^{-j2\pi {{\varphi }_{{{m}'}}}{{t}_{l}}}} \right\}} \\ 
& =\sum\limits_{m=1}^{{{N}_{T}}}{\left\{ \mu _{m,n}^{{{m}'}}\left( \tau ;{{r}_{t}},{{\psi }_{t}} \right){{e}^{j2\pi \left[ {{D}_{m}}-{{D}_{{{m}'}}} \right]{{t}_{l}}}} \right\}} \\ 
& =\mu _{{m}',n}^{{{m}'}}\left( \tau ;{{r}_{t}},{{\psi }_{t}} \right){{e}^{j2\pi \frac{{{f}_{{{m}'}}}}{{{f}_{c}}}{{f}_{d}}{{t}_{l}}}} \\ 
&\kern 9pt +\underbrace{\sum\limits_{\begin{smallmatrix} 
		m=1 \\ 
		m\ne {m}' 
		\end{smallmatrix}}^{{{N}_{T}}}{\left\{ \mu _{m,n}^{{{m}'}}\left( \tau ;{{r}_{t}},{{\psi }_{t}} \right){{e}^{j2\pi \left[ {{D}_{m}}-{{D}_{{{m}'}}} \right]{{t}_{l}}}} \right\}}}_{\text{aliased \kern 2pt term}} \\ 
\end{aligned}
\end{equation}
where
\begin{subequations}
	\begin{equation}
\kern -40pt \mu _{m,n}^{{{m}'}}\left( \tau ;{{r}_{t}},{{\psi }_{t}} \right)={{\delta }_{t}}{{a}_{m,n}}\left( \xi \left( {{r}_{t}} \right),{{\psi }_{t}} \right)g_{m}^{{{m}'}}\left( \tau;w_m  \right)
	\end{equation}
	\begin{equation}
\kern 47.2pt {{D}_{m}}=\frac{{{f}_{c}}+\left( m-1 \right)\Delta f}{{{f}_{c}}}{{f}_{t,d}}+m\Delta f+{{\varphi }_{m}}
	\end{equation}
\end{subequations}
Here, the symbol $\hat{t_l}$ denotes the processing in the slow-time domain.
\newcounter{TempEqCnt2} % 创建临时变量TempEqCnt
\setcounter{TempEqCnt2}{\value{equation}} % 将当前公式序号 赋给TempEqCnt
\setcounter{equation}{37} % 当前公式序号变为x，x等于长公式应有的序号减1.
\begin{figure*}[ht] %hb代表放在文章底部，%ht为放在文章顶部
	\hrulefill
	\begin{subequations}
	\begin{equation}
\kern -32pt \left\{ {{\varphi }_{m}}\left| \begin{matrix}
{{D}_{m+1}}-{{D}_{m}}\ge \frac{\text{PRF}}{{{N}_{T}}}  \\
{{D}_{{{N}_{T}}}}\le \text{PRF}  \\
\end{matrix} \right. \right\}=\left\{ {{\varphi }_{m}}\left| \begin{matrix}
\frac{\Delta f}{{{f}_{c}}}{{f}_{t,d}}+\Delta f+{{\varphi }_{m+1}}-{{\varphi }_{m}}\ge \frac{\text{PRF}}{{{N}_{T}}}  \\
\frac{{{f}_{{{N}_{T}}}}}{{{f}_{c}}}{{f}_{t,d}}+{{N}_{T}}\Delta f+{{\varphi }_{{{N}_{T}}}}\le \text{PRF}  \\
\end{matrix} \right. \right\}\
\end{equation}
		\begin{equation}
\kern 45pt \left\{ {{\varphi }_{m}}\left| \begin{matrix}
{{D}_{m+1}}-{{D}_{m}}\ge \frac{\text{PRF}}{{{N}_{T}}}  \\
{{D}_{{{N}_{T}}}}\ge \text{PRF}  \\
{{D}_{1}}-\left( {{D}_{{{N}_{T}}}}-\text{PRF} \right)\ge \frac{\text{PRF}}{{{N}_{T}}}  \\
\end{matrix} \right. \right\}=\left\{ {{\varphi }_{m}}\left| \begin{matrix}
\frac{\Delta f}{{{f}_{c}}}{{f}_{t,d}}+\Delta f+{{\varphi }_{m+1}}-{{\varphi }_{m}}\ge \frac{\text{PRF}}{{{N}_{T}}}  \\
\frac{{{f}_{{{N}_{T}}}}}{{{f}_{c}}}{{f}_{t,d}}+{{N}_{T}}\Delta f+{{\varphi }_{{{N}_{T}}}}\ge \text{PRF}  \\
\frac{\left( 1-{{N}_{T}} \right)\Delta f}{{{f}_{c}}}{{f}_{t,d}}+\left( 1-{{N}_{T}} \right)\Delta f+{{\varphi }_{1}}-{{\varphi }_{m+1}}\text{+PRF}\ge \frac{\text{PRF}}{{{N}_{T}}}  \\
\end{matrix} \right. \right\}
		\end{equation}
	\end{subequations}
\end{figure*}

\subsection{Low-pass Filtering}
Note that by designing the phase code ${{\varphi }_{m}}$, it is possible to make the Doppler of the aliased term and the Doppler of the desired signal $\mu _{{m}',n}^{{{m}'}}\left( \tau ;{{r}_{t}},{{\psi }_{t}} \right){{e}^{j2\pi \frac{{{f}_{{{m}'}}}}{{{f}_{c}}}{{f}_{d}}{{t}_{l}}}}$ belong to different Doppler bands.
Accordingly, under the condition that the target Doppler ${{f}_{t,d}}$ is smaller than $\frac{\text{PRF}}{{{N}_{T}}}$, where $\text{PRF}$ denotes the pulse repetition frequency, the constraints that need to be imposed on the phase code ${{\varphi }_{m}}$ are given in $(38)$, shown at the top of the next page, where the $N_T$ different Doppler frequencies, $\left\{ {{D}_{m}} \right\}_{m=1}^{{{N}_{T}}}$, given in $(37b)$, are forced to lie in the unambiguous Doppler domain with a mutual Doppler gap of $\frac{\text{PRF}}{{{N}_{T}}}$.
As a result, applying a low-pass filter with cutoff frequency $\frac{\text{PRF}}{{{N}_{T}}}$ in the Doppler domain, the aliased term can be filtered out.

Thus, sampling the signal ${{\tilde{r}}_{{m}',n}}\left( {{{\hat{t}}}_{l}},\tau  \right)$ at $\tau =\xi \left( {{r}_{t}} \right)$ and then stacking the $L$ outputs obtained by the low-pass filtering yields
\begin{equation}
\begin{aligned}
 {{{\mathbf{\overset{\lower0.5em\hbox{$\smash{\scriptscriptstyle\smile}$}}{r}}}}_{{m}',n}}\left( {{r}_{t}} \right)&={{{\mathbf{\tilde{r}}}}_{{m}',n}}\left( {{t}_{l}},{{r}_{t}} \right)*\mathbf{\bar{h}}\left( {{t}_{l}} \right)\\ 
& =\mu _{{m}',n}^{{{m}'}}\left( {{r}_{t}} \right)\mathbf{b}_{dop}^{{{m}'}}\left( {{f}_{t,d}} \right) \\ 
\end{aligned}
\end{equation}
where
\begin{subequations}
	\begin{equation}
	\begin{aligned}
	& {{{\mathbf{\tilde{r}}}}_{{m}',n}}\left( {{t}_{l}},{{r}_{t}} \right)= \\ 
	& {{\left[ {{{\tilde{r}}}_{{m}',n}}\left( {{t}_{1}},{{r}_{t}} \right),{{{\tilde{r}}}_{{m}',n}}\left( {{t}_{2}},{{r}_{t}} \right),...,{{{\tilde{r}}}_{{m}',n}}\left( {{t}_{L}},{{r}_{t}} \right) \right]}^{T}} \\ 
	\end{aligned}
	\end{equation}
	\begin{equation}
	\kern -60pt {{{\tilde{r}}}_{{m}',n}}\left( {{t}_{1}},{{r}_{t}} \right)={{{\tilde{r}}}_{{m}',n}}\left( {{t}_{1}},\tau =\xi \left( {{r}_{t}} \right) \right),
	\end{equation}
\end{subequations}
	\begin{equation}
\kern 23pt \begin{aligned}
\mu _{{m}',n}^{{{m}'}}\left( {{r}_{t}} \right)&=\mu _{{m}',n}^{{{m}'}}\left( \tau =\xi \left( {{r}_{t}} \right);{{r}_{t}},{{\psi }_{t}} \right) \\ 
& ={{\delta }_{t}}{{a}_{m,n}}\left( \xi \left( {{r}_{t}} \right),{{\psi }_{t}} \right)g_{{{m}'}}^{{{m}'}}\left( \tau =\xi \left( {{r}_{t}} \right);w_m \right) \\ 
& ={{\delta }_{t}}{{a}_{m,n}}\left( \xi \left( {{r}_{t}} \right),{{\psi }_{t}} \right){{w}_{m}}\mathsf{\mathbb{G}}\left( 0,0 \right), \\ 
\end{aligned}
\end{equation}
\begin{equation}
\kern -36pt \mathbf{b}_{dop}^{{{m}'}}\left( {{f}_{t,d}} \right)=\left[ \begin{matrix}
{{e}^{j2\pi \frac{{{f}_{c}}+\left( {m}'-1 \right)\Delta f}{{{f}_{c}}}{{f}_{t,d}}{{t}_{1}}}}  \\
{{e}^{j2\pi \frac{{{f}_{c}}+\left( {m}'-1 \right)\Delta f}{{{f}_{c}}}{{f}_{t,d}}{{t}_{2}}}}  \\
...  \\
{{e}^{j2\pi \frac{{{f}_{c}}+\left( {m}'-1 \right)\Delta f}{{{f}_{c}}}{{f}_{t,d}}{{t}_{L}}}}  \\
\end{matrix} \right],
\end{equation}
with $\mathbf{\bar{h}}\left( {{t}_{l}} \right)\in {{\mathsf{\mathbb{C}}}^{L\times 1}}$ denoting the low-pass filter function in the Doppler domain.
With $N_R$ receive antennas, each of which has $N_T$ channels, the received data for one CPI comprises $N_TN_RL$ complex samples.
Thus, the ${{N}_{T}}{{N}_{R}}L\times 1$-dimensional space-time snapshot can be obtained as
\begin{equation}
\begin{aligned}
& {{\mathbf{q}}_{tar}}\left( {{r}_{t}},{{\psi }_{t}},{{f}_{t,d}};\mathbf{w} \right)={{\delta }_{t}}\mathbb{G}\left( 0,0 \right) \\ 
& \times \operatorname{vec}\left\{ \left[ \begin{matrix}
{{\mathbf{k}}_{1}}\left( \mathbf{w} \right) & {{\mathbf{k}}_{2}}\left( \mathbf{w} \right) & ... & {{\mathbf{k}}_{L}}\left( \mathbf{w} \right)  \\
\end{matrix} \right] \right\} \\ 
\end{aligned}
\end{equation}
where
\begin{subequations}
	\begin{equation}
\kern 10pt {{\mathbf{k}}_{l}\left( \mathbf{w} \right)}=\left( {{\mathbf{I}}_{{{N}_{R}}}}\otimes \operatorname{diag}\left\{ \mathbf{w}\odot {{{\mathbf{\bar{b}}}}_{dop,l}} \right\} \right)\left[ \begin{matrix}
{{\mathbf{a}}_{1}}\left( \xi \left( {{r}_{t}} \right),{{\psi }_{t}} \right)  \\
{{\mathbf{a}}_{2}}\left( \xi \left( {{r}_{t}} \right),{{\psi }_{t}} \right)  \\
...  \\
{{\mathbf{a}}_{{{N}_{R}}}}\left( \xi \left( {{r}_{t}} \right),{{\psi }_{t}} \right)  \\
\end{matrix} \right]
	\end{equation}
	\begin{equation}
\kern -80pt {{{\mathbf{\bar{b}}}}_{dop,l}}=\left[ \begin{matrix}
{{e}^{j2\pi {{f}_{t,d}}{{t}_{l}}}}  \\
{{e}^{j2\pi \frac{{{f}_{c}}+\Delta f}{{{f}_{c}}}{{f}_{t,d}}{{t}_{l}}}}  \\
...  \\
{{e}^{j2\pi \frac{{{f}_{c}}+\left( {{N}_{T}}-1 \right)\Delta f}{{{f}_{c}}}{{f}_{t,d}}{{t}_{l}}}}  \\
\end{matrix} \right]
	\end{equation}
\end{subequations}
for $l=1,2,...L$.
Considering that $\Delta f\ll {{f}_{c}}$
\begin{equation}
\begin{aligned}
{{a}_{m,n}}\left( {{\psi }_{t}} \right)&={{e}^{j2\pi \left( {{f}_{c}}+\left( m-1 \right)\Delta f \right)\left[ \left( m-1 \right){{\xi }_{T}}\left( {{\psi }_{t}} \right)+\left( n-1 \right){{\xi }_{R}}\left( {{\psi }_{t}} \right) \right]}} \\ 
& \approx {{e}^{j2\pi \left( {{f}_{c}}+\left( m-1 \right)\Delta f \right)\left( m-1 \right){{\xi }_{T}}\left( {{\psi }_{t}} \right)}}{{e}^{j2\pi {{f}_{c}}\left( n-1 \right){{\xi }_{R}}\left( {{\psi }_{t}} \right)}}, \\ 
\end{aligned}
\end{equation}
allowing $(43)$ to be approximated as
\begin{equation}
\begin{aligned}
& {{\mathbf{q}}_{tar}}\left( {{r}_{t}},{{\psi }_{t}},{{f}_{t,d}};\mathbf{w} \right) \\ 
& \approx {{\delta }_{t}}\mathsf{\mathbb{G}}\left( 0,0 \right){{\mathbf{b}}_{dop}}\left( {{f}_{t,d}} \right)\otimes \left( {{\mathbf{a}}_{R}}\left( {{\psi }_{t}} \right)\otimes \left[ \mathbf{w}\odot {{\mathbf{a}}_{T}}\left( {{r}_{t}},{{\psi }_{t}} \right) \right] \right) \\ 
& ={{\alpha }_{t}}\mathbf{\bar{q}}\left( {{r}_{t}},{{\psi }_{t}},{{f}_{t,d}} ;\mathbf{w}\right) \\ 
\end{aligned}
\end{equation}
where
$\otimes $ denotes the Kronecker matrix product,
\begin{subequations}
	\begin{equation}
	\kern -96pt {{\alpha }_{t}}={{\delta }_{t}}\mathsf{\mathbb{G}}\left( 0,0 \right)
	\end{equation}
	\begin{equation}
	{{\mathbf{b}}_{dop}}\left( {{f}_{t,d}} \right)={{\left[ \begin{matrix}
			{{e}^{j2\pi {{f}_{t,d}}{{t}_{1}}}} & {{e}^{j2\pi {{f}_{t,d}}{{t}_{2}}}} & ... & {{e}^{j2\pi {{f}_{t,d}}{{t}_{L}}}}  \\
			\end{matrix} \right]}^{T}}
	\end{equation}
\end{subequations}
and 
\begin{equation}
\mathbf{\bar{q}}\left( r,\psi ,{{f}_{d}};\mathbf{w} \right)={{\mathbf{b}}_{dop}}\left( {{f}_{d}} \right)\otimes \left( {{\mathbf{a}}_{R}}\left( \psi  \right)\otimes \left[ \mathbf{w}\odot {{\mathbf{a}}_{T}}\left( r,\psi  \right) \right] \right)
\end{equation}
where
\begin{subequations}
\begin{equation}
\kern -67pt {{\mathbf{a}}_{T}}\left( {{r}_{t}},{{\psi }_{t}} \right)={{\mathbf{a}}_{T}}\left( {{\psi }_{t}} \right)\odot \mathbf{e}\left( \xi \left( {{r}_{t}} \right) \right)
\end{equation}
\begin{equation}
\kern 32pt{{\mathbf{a}}_{T}}\left( {{\psi }_{t}} \right)=\left[ \begin{matrix}
1  \\
{{e}^{j2\pi \left( {{f}_{c}}+\Delta f \right){{\xi }_{T}}\left( {{\psi }_{t}} \right)}}  \\
...  \\
{{e}^{j2\pi \left( {{f}_{c}}+\left( {{N}_{T}}-1 \right)\Delta f \right)\left( {{N}_{T}}-1 \right){{\xi }_{T}}\left( {{\psi }_{t}} \right)}}  \\
\end{matrix} \right]
\end{equation}
		\begin{equation}
\kern -42pt {{\mathbf{a}}_{R}}\left( {{\psi }_{t}} \right)=\left[ \begin{matrix}
1  \\
{{e}^{j2\pi {{f}_{c}}{{\xi }_{R}}\left( {{\psi }_{t}} \right)}}  \\
...  \\
{{e}^{j2\pi {{f}_{c}}\left( {{N}_{R}}-1 \right){{\xi }_{R}}\left( {{\psi }_{t}} \right)}}  \\
\end{matrix} \right]
	\end{equation}
\end{subequations}
Fig. 3 shows the resulting proposed receiver for a coherent airborne FDA.
For each receive antenna, a joint space-time processing module is applied to extract the desired signals from the aliased returns.

\begin{figure}[t]
	\centering  %图片全局居中
	\subfigure[The receiver structure.]{
		\label{Fig.sub.33}
		\includegraphics[width=0.42\textwidth]{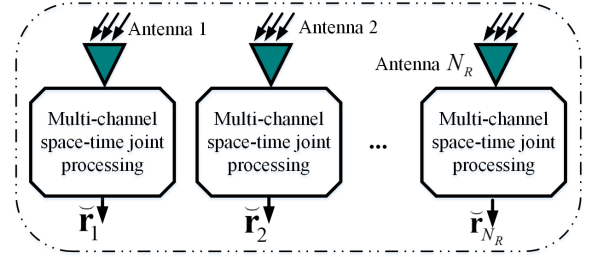}}
	\subfigure[The joint space-time processing module.]{
		\label{Fig.sub.44}
		\includegraphics[width=0.42\textwidth]{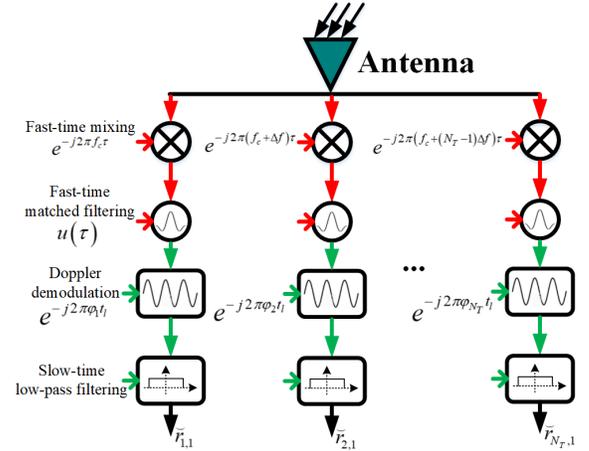}}
	\caption{The proposed FDA receiver based on joint space-time processing.}
	\label{Fig.main22}
\end{figure}
\section{FDA Range-space-time processing}
\subsection{Clutter Signal Model}
For an airborne surveillance radar, the Earth's surface is the major source of clutter, which further complicates processing by the fact that the return from a discrete ground clutter source has the same form as a target echo. Since the ground is assumed to have zero inherent velocity, the relative Doppler of a ground clutter source depends only on its aspect with respect to the radar and on the platform velocity.
Therefore, the clutter signal from the $j$-th clutter patch of range cell $r_{c,i}$ can be expressed as
\begin{equation}
{{\mathbf{q}}_{clu,i,j}\left(\mathbf{w}\right)}={{\alpha }_{c,i,j}}\mathbf{\bar{q}}\left( {{r}_{c,i}},{{\psi }_{c,j}},{{f}_{c,d,j}} ;\mathbf{w}\right),
\end{equation}
for $j=1,2,...,J$, with $J$ denoting the number of range cells, and
$i=1,2,...,I$, with $I$ denoting the number of clutter patches in an iso-range ring, where
\begin{equation}
{{f}_{d,c,j}}=\frac{2{{v}_{a}}}{\lambda }\cos {{\psi }_{c,j}}
\end{equation}
and where ${{\alpha }_{c,i,j}}$ denotes the complex-valued clutter amplitude. 

\subsection{Jamming Signal Model}
In this work, we only consider barrage jamming that is spatially correlated from element to element and temporally uncorrelated from pulse to pulse.
Thus, the barrage jamming will temporally appear as thermal noise, but as a point target or a discrete clutter source in the spatial domain.
The space-time snapshot of ${\tilde{j}}$-th jamming may be written as
\begin{equation}
{{\mathbf{q}}_{jam,\tilde{j}}}={{\alpha }_{jam,\tilde{j}}}\mathbf{\bar{u}}\otimes {{\mathbf{a}}_{R}}\left( {{\psi }_{jam,\tilde{j}}} \right)\otimes \mathbf{\tilde{u}}
\end{equation}
for $\tilde{j}=1,2,...,\tilde{J}$, with $\tilde{J}$ denoting the number of barrage jammers,
where $\mathbf{\bar{u}}\in {{\mathsf{\mathbb{C}}}^{L\times 1}}$ and $\mathbf{\tilde{u}}\in {{\mathsf{\mathbb{C}}}^{{{N}_{T}}\times 1}}$ are Gaussian random vectors.

\subsection{Range-space-time Processing}
Taking into account the clutter, jamming, and receiver noise, the coherent FDA receive data model may expressed as
\begin{equation}
\kern -2pt \begin{aligned}
\mathbf{q}&={{\mathbf{q}}_{tar}}\left( {{r}_{t}},{{\psi }_{t}},{{f}_{t,d}};\mathbf{w} \right) \\ 
& \kern 10pt +\sum\limits_{i}^{I}{\sum\limits_{j}^{J}{{{\mathbf{q}}_{clu,i,j}}\left( \mathbf{w} \right)}}+\sum\limits_{{\tilde{j}}}^{{\tilde{J}}}{{{\mathbf{q}}_{jam,\tilde{j}}}}+\mathbf{n} \\ 
& ={{\alpha }_{t}}\mathbf{\bar{q}}\left( {{r}_{t}},{{\psi }_{t}},{{f}_{t,d}};\mathbf{w} \right)+\sum\limits_{i}^{I}{\sum\limits_{j}^{J}{{{\alpha }_{c,i,j}}\mathbf{\bar{q}}\left( {{r}_{c,i}},{{\psi }_{c,j}},{{f}_{c,d,j}};\mathbf{w} \right)}} \\ 
&\kern 10pt +\sum\limits_{{\tilde{j}}}^{{\tilde{J}}}{{{\alpha }_{jam,\tilde{j}}}\mathbf{\bar{u}}\otimes {{\mathbf{a}}_{R}}\left( {{\psi }_{jam,j}} \right)\otimes \mathbf{\tilde{u}}}+\mathbf{n}. \\ 
\end{aligned}
\end{equation}
where $\mathbf{n}\in {{\mathsf{\mathbb{C}}}^{{{N}_{T}}{{N}_{R}}L\times 1}}$ denotes an additive circularly symmetric Gaussian noise with power ${{\left| {{\alpha }_{noise}} \right|}^{2}}$.
Employing a receiving beamformer $\mathbf{v}\in {{\mathsf{\mathbb{C}}}^{{{N}_{T}}{{N}_{R}}L\times 1}}$ to synthesis the multiple channel outputs yields the output signal-to-interference-plus-noise ratio (SINR) 
\begin{equation}
\begin{aligned}
&\kern -4pt \psi \left( \mathbf{v},\mathbf{w} \right)\\
&=\frac{\mathbb{E}\left\{ {{\left| {{\mathbf{v}}^{H}}{{\mathbf{q}}_{tar}}\left( {{r}_{t}},{{\psi }_{t}},{{f}_{t,d}};\mathbf{w} \right) \right|}^{2}} \right\}}{\mathbb{E}\left\{ {{\left| {{\mathbf{v}}^{H}}\left( \mathbf{q}-{{\mathbf{q}}_{tar}}\left( {{r}_{t}},{{\psi }_{t}},{{f}_{t,d}};\mathbf{w} \right) \right) \right|}^{2}} \right\}} \\ 
& =\frac{{{\left| {\mathbb{E}\left\{ {{\left| {{\alpha }_{t}} \right|}^{2}} \right\}} \right|}^{2}}{{\mathbf{v}}^{H}}\mathbf{\bar{q}}\left( {{r}_{t}},{{\psi }_{t}},{{f}_{t,d}};\mathbf{w} \right){{{\mathbf{\bar{q}}}}^{H}}\left( {{r}_{t}},{{\psi }_{t}},{{f}_{t,d}};\mathbf{w} \right)\mathbf{v}}{{{\mathbf{v}}^{H}}{{\mathbf{R}}_{cjn}}\left( \mathbf{w} \right)\mathbf{v}} \\ 
& =\text{SNR}\cdot \frac{{{\mathbf{v}}^{H}}\mathbf{\bar{q}}\left( {{r}_{t}},{{\psi }_{t}},{{f}_{t,d}};\mathbf{w} \right){{{\mathbf{\bar{q}}}}^{H}}\left( {{r}_{t}},{{\psi }_{t}},{{f}_{t,d}};\mathbf{w} \right)\mathbf{v}}{{{\mathbf{v}}^{H}}{{{\mathbf{\bar{R}}}}_{cjn}}\left( \mathbf{w} \right)\mathbf{v}} \\ 
\end{aligned}
\end{equation}
where $\mathbb{E}\left\{ \cdot  \right\}$ is the expectation operator and ${{\mathbf{R}}_{cjn}}$ denotes the interference-plus-noise covariance matrix, given by
\begin{equation}
\begin{aligned}
{{\mathbf{R}}_{cjn}}\left( \mathbf{w} \right)&={{\mathbf{R}}_{clu}}\left( \mathbf{w} \right)+{{\mathbf{R}}_{jam}}+{{\mathbf{R}}_{noise}} \\ 
& ={{\left| {{\alpha }_{noise}} \right|}^{2}}{{{\mathbf{\bar{R}}}}_{cjn}}\left( \mathbf{w} \right) \\ 
\end{aligned}
\end{equation}
where
\begin{equation}
\begin{aligned}
& {{\mathbf{R}}_{clu}}\left( \mathbf{w} \right) \\ 
& =\mathbb{E}\left\{ \left( \sum\limits_{i}^{I}{\sum\limits_{j}^{J}{{{\mathbf{q}}_{clu,i,j}}\left( \mathbf{w} \right)}} \right){{\left( \sum\limits_{i}^{I}{\sum\limits_{j}^{J}{{{\mathbf{q}}_{clu,i,j}}\left( \mathbf{w} \right)}} \right)}^{H}} \right\} \\ 
& ={\mathbb{E}\left\{ {{\left| {{\alpha }_{c,i,j}} \right|}^{2}} \right\}} \\ 
& \kern 11pt \times \sum\limits_{i}^{I}{\sum\limits_{j}^{J}{\mathbf{\bar{q}}\left( {{r}_{c,i}},{{\psi }_{c,j}},{{f}_{c,d,j}};\mathbf{w} \right){{{\mathbf{\bar{q}}}}^{H}}\left( {{r}_{c,i}},{{\psi }_{c,j}},{{f}_{c,d,j}};\mathbf{w} \right)}} \\ 
\end{aligned}
\end{equation}
denotes the clutter covariance matrix. Furthermore,
	\begin{equation}
\begin{aligned}
{{\mathbf{R}}_{jam}}&=\mathbb{E}\left\{ \left( \sum\limits_{{\tilde{j}}}^{{\tilde{J}}}{{{\mathbf{q}}_{jam,\tilde{j}}}} \right){{\left( \sum\limits_{{\tilde{j}}}^{{\tilde{J}}}{{{\mathbf{q}}_{jam,\tilde{j}}}} \right)}^{H}} \right\} \\ 
& ={{\mathbf{I}}_{L}}\otimes \left[ \sum\limits_{{\tilde{j}}}^{{\tilde{J}}}{{\mathbb{E}\left\{ {{\left| {{\alpha }_{{jam,\tilde{j}}}} \right|}^{2}} \right\}}{{\mathbf{a}}_{R}}\left( {{\theta }_{{\tilde{j}}}} \right)\mathbf{a}_{R}^{H}\left( {{\theta }_{{\tilde{j}}}} \right)} \right]\otimes {{\mathbf{I}}_{N_T}} \\ 
\end{aligned}
\end{equation}
denotes the jammer covariance matrix, with ${{\mathbf{I}}_{K}}$ being the $K\times K$-dimensional identity matrix,
\begin{equation}
{{\mathbf{R}}_{noise}}=\mathbb{E}\left\{ \mathbf{n}{{\mathbf{n}}^{H}} \right\}={{\left| {{\alpha }_{noise}} \right|}^{2}}{{\mathbf{I}}_{N_TN_RL}}
\end{equation}
denotes the noise covariance matrix, and
\begin{equation}
\begin{aligned}
& {{{\mathbf{\bar{R}}}}_{cjn}}\left( \mathbf{w} \right)= \\ 
& \sum\limits_{i}^{I}{\sum\limits_{j}^{J}{\text{CN}{{\text{R}}_{i,j}}\mathbf{\bar{q}}\left( {{r}_{c,i}},{{\psi }_{c,j}},{{f}_{c,d,j}};\mathbf{w} \right){{{\mathbf{\bar{q}}}}^{H}}\left( {{r}_{c,i}},{{\psi }_{c,j}},{{f}_{c,d,j}};\mathbf{w} \right)}} \\ 
& \kern 16pt +{{\mathbf{I}}_{L}}\otimes \left[ \sum\limits_{{\tilde{j}}}^{{\tilde{J}}}{\text{JN}{{\text{R}}_{{\tilde{j}}}}{{\mathbf{a}}_{R}}\left( {{\theta }_{{\tilde{j}}}} \right)\mathbf{a}_{R}^{H}\left( {{\theta }_{{\tilde{j}}}} \right)} \right]\otimes {{\mathbf{I}}_{{{N}_{T}}}}+{{\mathbf{I}}_{{{N}_{T}}{{N}_{R}}L}} \\ 
\end{aligned}
\end{equation}
denotes the interference plus noise covariance matrix.
Here,
$\text{SNR}=\frac{\mathbb{E}\left\{ {{\left| {{\alpha }_{t}} \right|}^{2}} \right\}}{{{\left| {{\alpha }_{noise}} \right|}^{2}}}$, $\text{CN}{{\text{R}}_{i,j}}=\frac{\mathbb{E}\left\{ {{\left| {{\alpha }_{c,i,j}} \right|}^{2}} \right\}}{{{\left| {{\alpha }_{noise}} \right|}^{2}}}$, and $\text{JN}{{\text{R}}_{{\tilde{j}}}}=\frac{\mathbb{E}\left\{ {{\left| {{\alpha }_{{jam,\tilde{j}}}} \right|}^{2}} \right\}}{{{\left| {{\alpha }_{noise}} \right|}^{2}}}$ denote the signal-to-noise ratio (SNR), the clutter-to-noise ratio (CNR), and the jamming-to-noise ratio (JNR), respectively.
Consequently, the output SINR maximization problem can be formulated as
\begin{equation}
\begin{matrix}
\underset{\mathbf{v},\mathbf{w}}{\mathop{\max }}\, & \frac{{{\mathbf{v}}^{H}}\mathbf{\bar{q}}\left( {{r}_{t}},{{\psi }_{t}},{{f}_{t,d}};\mathbf{w} \right){{{\mathbf{\bar{q}}}}^{H}}\left( {{r}_{t}},{{\psi }_{t}},{{f}_{t,d}};\mathbf{w} \right)\mathbf{v}}{{{\mathbf{v}}^{H}}{{{\mathbf{\bar{R}}}}_{cjn}}\left( \mathbf{w} \right)\mathbf{v}}.  \\
\end{matrix}.
\end{equation}

In practical applications, the optimization variables are often subject to many constraints.
For example, constant modulus waveforms are often required due to the limitations of nonlinear radar amplifiers \cite{6649991,7450660}. 
Therefore, the above problem will typically also include multiple nonconvex constraints, making it a NP-hard problem which may not be solved efficiently  \cite{boyd2004convex}.
However, it may be noted that the SINR can also be expressed as an explicit expression of the transmit weight vector, $\mathbf{w}$, as shown in the Appendix.
This implies that an iterative optimization method based on a semidefinite relaxation (SDR) technique can be applied to solve it\footnote{How to solve the resulting non-convex problem more efficiently is beyond the scope of this paper, and interested readers are referred to \cite{cheng2019joint} for a further discussion on these aspects.} \cite{5447068}.
In this work, we consider uniform  transmit weights, i.e., $\mathbf{w}={{\mathbf{1}}_{{{N}_{T}}\times 1}}$, with ${{\mathbf{1}}_{{{N}_{T}}\times 1}}$ denoting the ${{N}_{T}}\times 1$-dimensional all-ones vector, converting $(60)$ to a minimum variance distortionless response (MVDR) problem, whose solution is given by (see, e.g., \cite{6649991,cheng2019joint})
\begin{equation}
\mathbf{v}=\frac{\mathbf{\bar{R}}_{cjn}^{-1}\left( \mathbf{w}={{\mathbf{1}}_{{{N}_{T}}\times 1}} \right){{{\mathbf{\bar{q}}}}_{\text{target},\mathbf{w}={{\mathbf{1}}_{{{N}_{T}}\times 1}}}}}{\mathbf{\bar{q}}_{\text{target},\mathbf{w}={{\mathbf{1}}_{{{N}_{T}}\times 1}}}^{H}\mathbf{\bar{R}}_{cjn}^{-1}\left( \mathbf{w}={{\mathbf{1}}_{{{N}_{T}}\times 1}} \right){{{\mathbf{\bar{q}}}}_{\text{target},\mathbf{w}={{\mathbf{1}}_{{{N}_{T}}\times 1}}}}}
\end{equation}
where ${{{\mathbf{\bar{q}}}}_{\text{target},\mathbf{w}={{\mathbf{1}}_{{{N}_{T}}\times 1}}}}=\mathbf{\bar{q}}\left( {{r}_{t}},{{\psi }_{t}},{{f}_{t,d}};\mathbf{w}={{\mathbf{1}}_{{{N}_{T}}\times 1}} \right)$.
As the covariance matrix ${{{\mathbf{\bar{R}}}}_{cjn}}\left( \mathbf{w} \right)$ for an FDA radar is range-dependent, it has to be estimated according to the actual scattering scenario, instead of using secondary data obtained from nearby range cells as is done in conventional radars \cite{wang2022covariance}.
As the estimation procedure is not the focus of this paper, the following simulations assume full knowledge of the covariance matrix.
It is worth noting that since the transmit weights provide more degrees of freedom to increase the SINR, the joint design in $(60)$ can be expected to obtain a higher SINR.

\section{Simulation}
In this section, numerical simulations are provided to evaluate the effectiveness of the proposed receiver and the performance of the joint range-space-time  processing.
Herein, we only consider clutter that is within the same range cell as the target. Unless otherwise specified, in all simulations, the  FDA system parameters are as listed in Tab. \uppercase\expandafter{\romannumeral1}.
Noting the relationship between the clutter Doppler and the platform velocity, given in $(51)$, the range of values for the clutter Doppler is $\left( -565,565 \right) \kern 2pt Hz$.
The parameters for the simulation scenarios are listed in Tab. \uppercase\expandafter{\romannumeral2}.

\begin{table*}
	\centering
	\caption{System parameters.}
	{\begin{tabular}[l]{@{}lcllcc}
			\toprule
			Parameter & Value & Parameter & Value\\
			\midrule
			Number of transmit antennas & $\kern -32pt {N_T}=5$  & Spacing of transmit array &  $\kern 5pt d_t = 0.125 \kern 2pt m$\\
			Number of receive antennas & \kern -37pt $ {N_R}=5$  & Spacing of receive array &  $\kern 5pt d_r = 0.125 \kern 2pt m$\\
			Pulse duration &\kern -20pt ${T_p}=1 \kern 2pt \mu s$  & Bandwidth of basedband LFM signal &\kern 10pt $20 \kern 2pt MHz$ \\
			Central carrier frequency &\kern 0pt ${f_c}=1.2 \kern 2pt GHz$  & Frequency offset & $\kern -2pt $ $\Delta f = 1 \kern 2.5pt MHz$\\
			Pulse repetition frequency &\kern -20pt ${\text{PRF}} = 7 \kern 2pt kHz$  & Wavelength &\kern 10pt $\lambda =0.25 \kern 2pt m$\\
			Number of pulses &\kern -20pt $L = 180$  & Platform velocity  &\kern 6pt ${{v}_{a}}=100 \kern 2pt {m}/{s}\;$\\
			\bottomrule
	\end{tabular}}
	\label{symbols}
\end{table*}

\begin{table*}
	\centering
	\caption{Parameter settings for simulation scenario.}
	{\begin{tabular}[l]{@{}lccccc}
			\toprule
			Type &Target  & Barrage jamming & Clutter\\
			\midrule
			Range $\left[ {km} \right]$  & $3.0$  & $3.0$ & $3.006$\\
			Azimuth angle & $45^o$  & $120^o$ & $\left( 0,{{180}^{o}} \right)$\\
			Depression angle & $45^o$  & $45^o$  & $45^o$\\
			Doppler $\left[ {Hz} \right]$ &$ 400$  & All Doppler  & $\left( -565,565 \right)$\\
			SNR/JNR/CNR $\left[ {dB} \right]$  &$0$  &$20$ & $20$\\
			\bottomrule
	\end{tabular}}
	\label{symbolss}
\end{table*}

\subsection{Adapted Patterns}
Fig. 4 shows the energy of the interferences distributed in the Doppler-azimuth domain, $P\left( {{f}_{d}},\psi  \right)$, as calculated by
\begin{equation}
\begin{aligned}
& P\left( \psi ,{{f}_{d}} \right)= \\ 
&\kern 6pt \frac{1}{{{{\mathbf{\bar{q}}}}^{H}}\left( {{r}_{t}},\psi ,{{f}_{d}} \right)\mathbf{\bar{R}}_{cjn}^{-1}\left( \mathbf{w} \right)\mathbf{\bar{q}}\left( {{r}_{t}},\psi ,{{f}_{d}};\mathbf{w} \right)}\left| _{\mathbf{w}={{\mathbf{1}}_{{{N}_{T}}\times 1}}} \right. \\ 
\end{aligned}.
\end{equation}
We refer the reader to  $(11)$ for the conversion of the conic angle and the azimuth angle.
As may be seen, the target is fully obscured by the clutter ridge.
\begin{figure}
	\centering
	\includegraphics[width=0.45\textwidth]{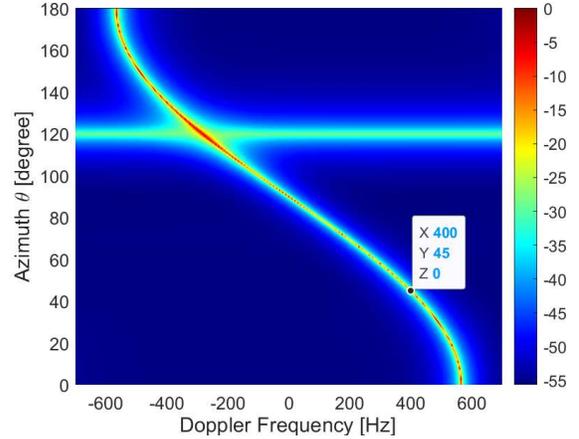}
	\caption{The energy distribution of the interferences in the Doppler-azimuth domain. The marked position is that of the target, which can be seen to be entirely obscured by the clutter ridge.}
\end{figure}
The adapted pattern, as a function of angle and Doppler, is an important performance metric, defined by 
\begin{equation}
\bar{P}\left( \psi ,{{f}_{d}} \right)={{\left| {{\mathbf{v}}^{H}}\mathbf{\bar{q}}\left( {{r}_{t}},\psi ,{{f}_{d}};\mathbf{w}={{\mathbf{1}}_{{{N}_{T}}\times 1}} \right) \right|}^{2}}.
\end{equation}
where for each angle-Doppler pair $\left( \psi ,{{f}_{d}} \right)$, the receive weight vector $\mathbf{v}$ is computed using $(61)$.
Ideally, the pattern has nulls in the locations of interference
sources and high gain at the angle and Doppler of the presumed target direction. 
Fig. 5 shows the adapted pattern behavior, with Figs. 6a and 6b showing the cuts of this pattern at the azimuth angle
$\theta ={{45}^{o}}$ and the Doppler frequency ${{f}_{d}}=400 \kern 2pt Hz$,
respectively.
One may note that the resulting pattern can form a mainlobe at the target location, with deep nulls at the locations of the jammer and clutter sources. 
In comparison, the results of the MIMO and the phased-array systems are shown in Figs. 7a and 7b, respectively, with the corresponding cuts shown in Fig. 8.
It is worth noting that for the MIMO and the phased-array systems, deep nulls appear along both the jamming and clutter lines.
However, there is no peak at the target location due to the lack of range-controllable degrees of freedom.

\begin{figure}
	\centering
	\includegraphics[width=0.45\textwidth]{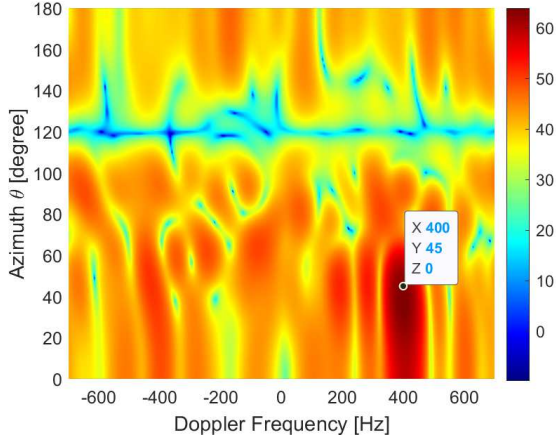}
	\caption{The output power distribution after implementing the joint range-space-time processing. The marked position is that of the target.}
\end{figure}

\begin{figure}[t]
	\centering  %图片全局居中
	\subfigure[]{
		\label{Fig.sub.32}
		\includegraphics[width=0.23\textwidth]{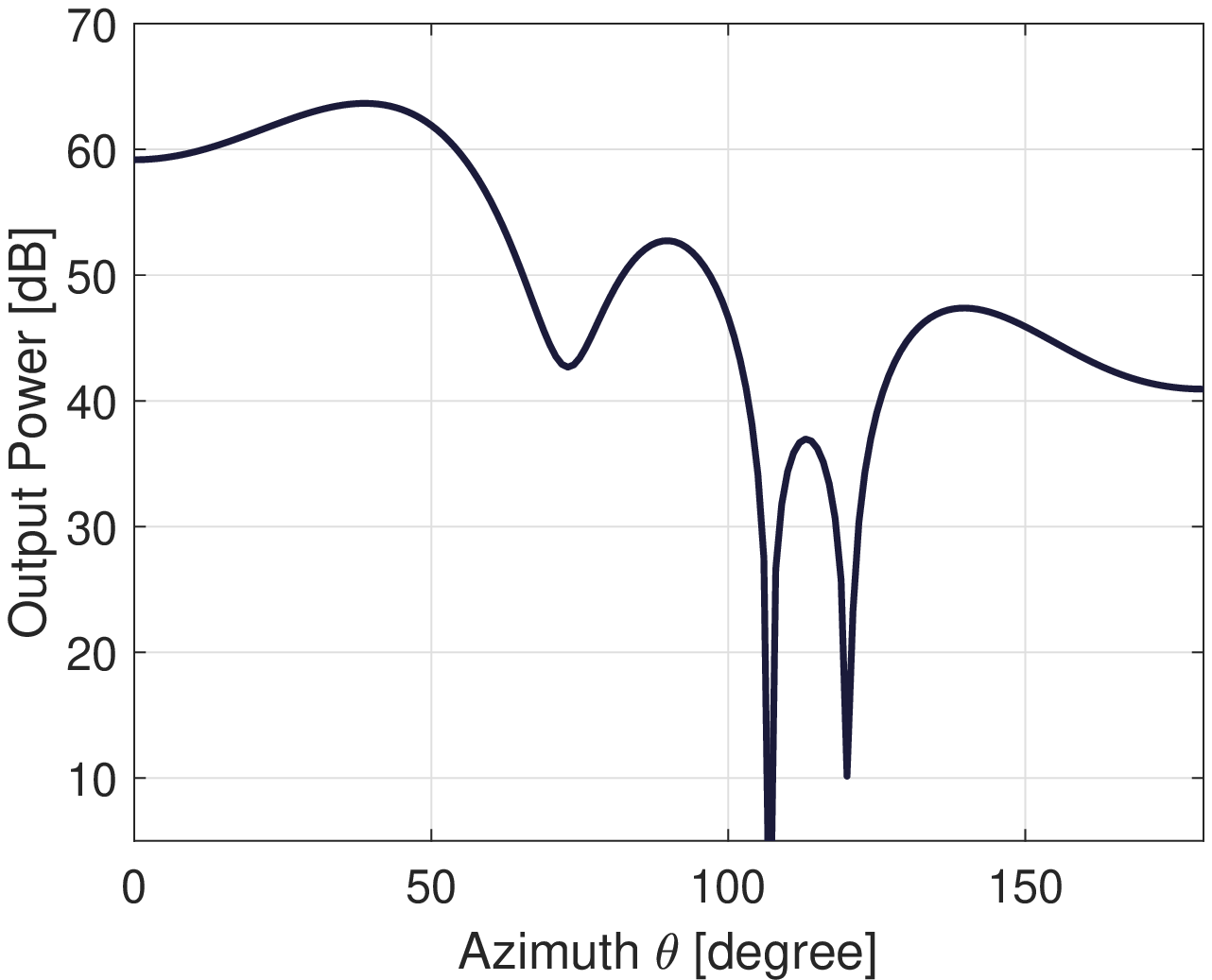}}
	\subfigure[]{
		\label{Fig.sub.41}
		\includegraphics[width=0.23\textwidth]{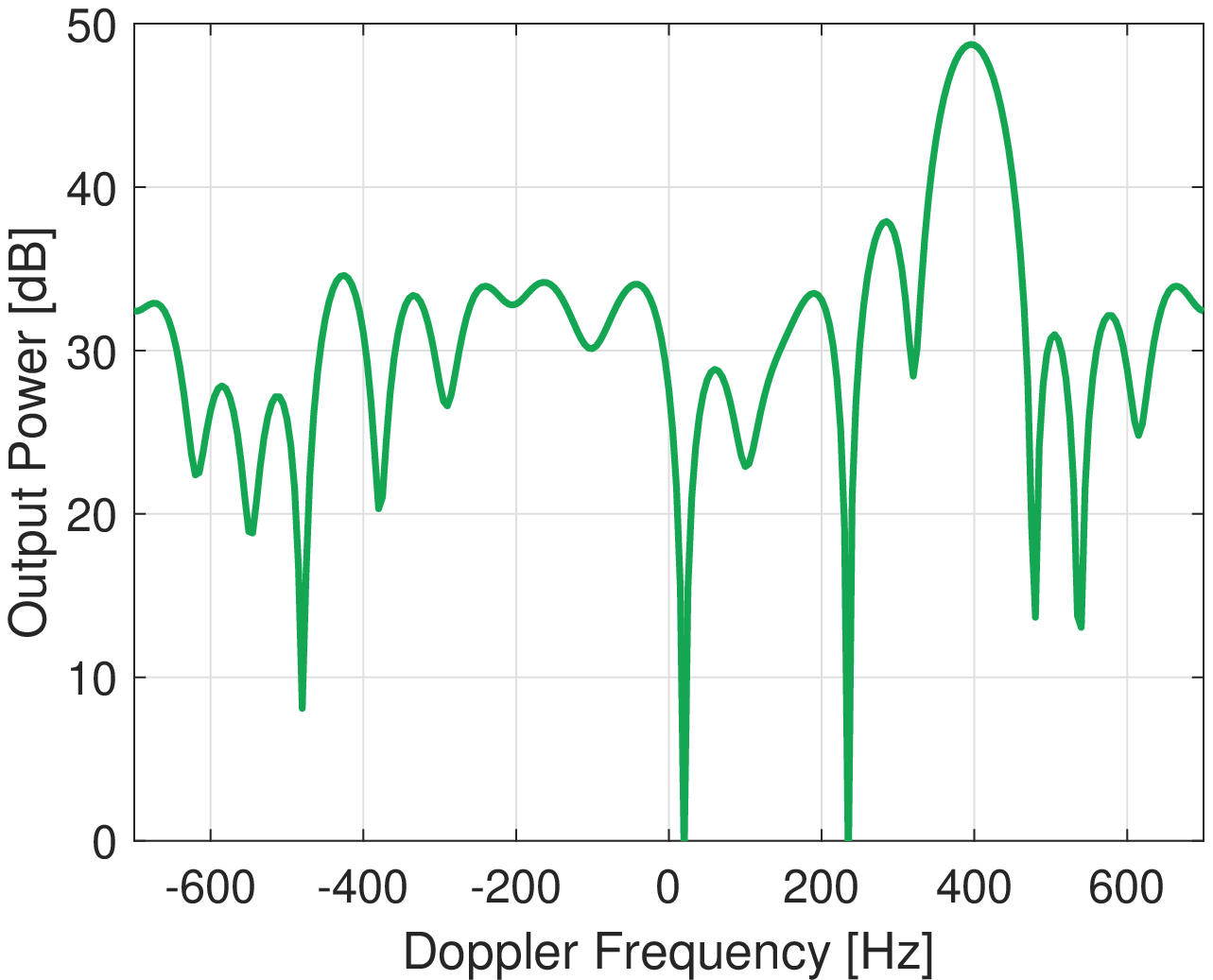}}
	\caption{Cuts of the FDA adapted pattern (a) at the Doppler frequency ${{f}_{d}}=400 \kern 2pt Hz$, and (b) at the azimuth angle
		$\theta ={{45}^{o}}$.  }
	\label{Fig.main2}
\end{figure}

\begin{figure}[t]
	\centering  %图片全局居中
	\subfigure[]{
		\label{Fig.sub.333}
		\includegraphics[width=0.40\textwidth]{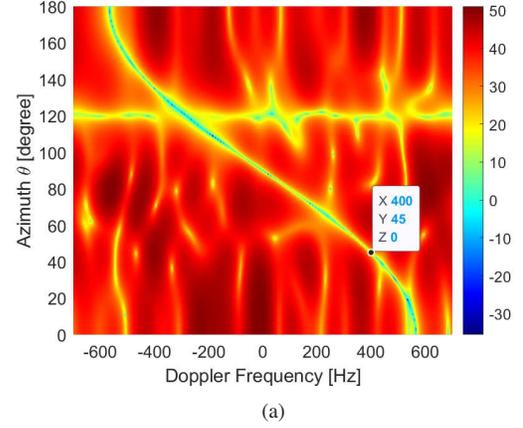}}
	\subfigure[]{
		\label{Fig.sub.433}
		\includegraphics[width=0.40\textwidth]{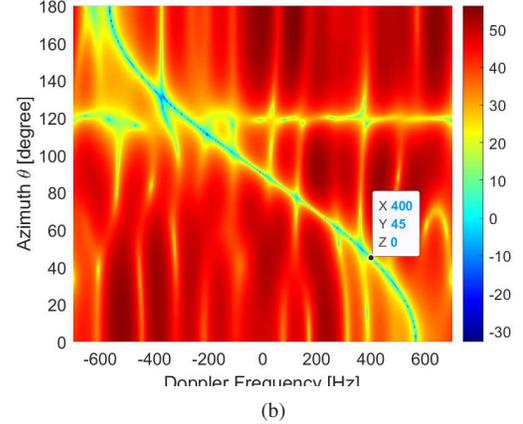}}
	\caption{The output power distribution of (a) a MIMO and (b) a phased-array system. The marked position is that of the target.}
	\label{Fig.main02}
\end{figure}

\begin{figure}[t]
	\centering  %图片全局居中
	\subfigure[]{
		\label{Fig.sub.3}
		\includegraphics[width=0.23\textwidth]{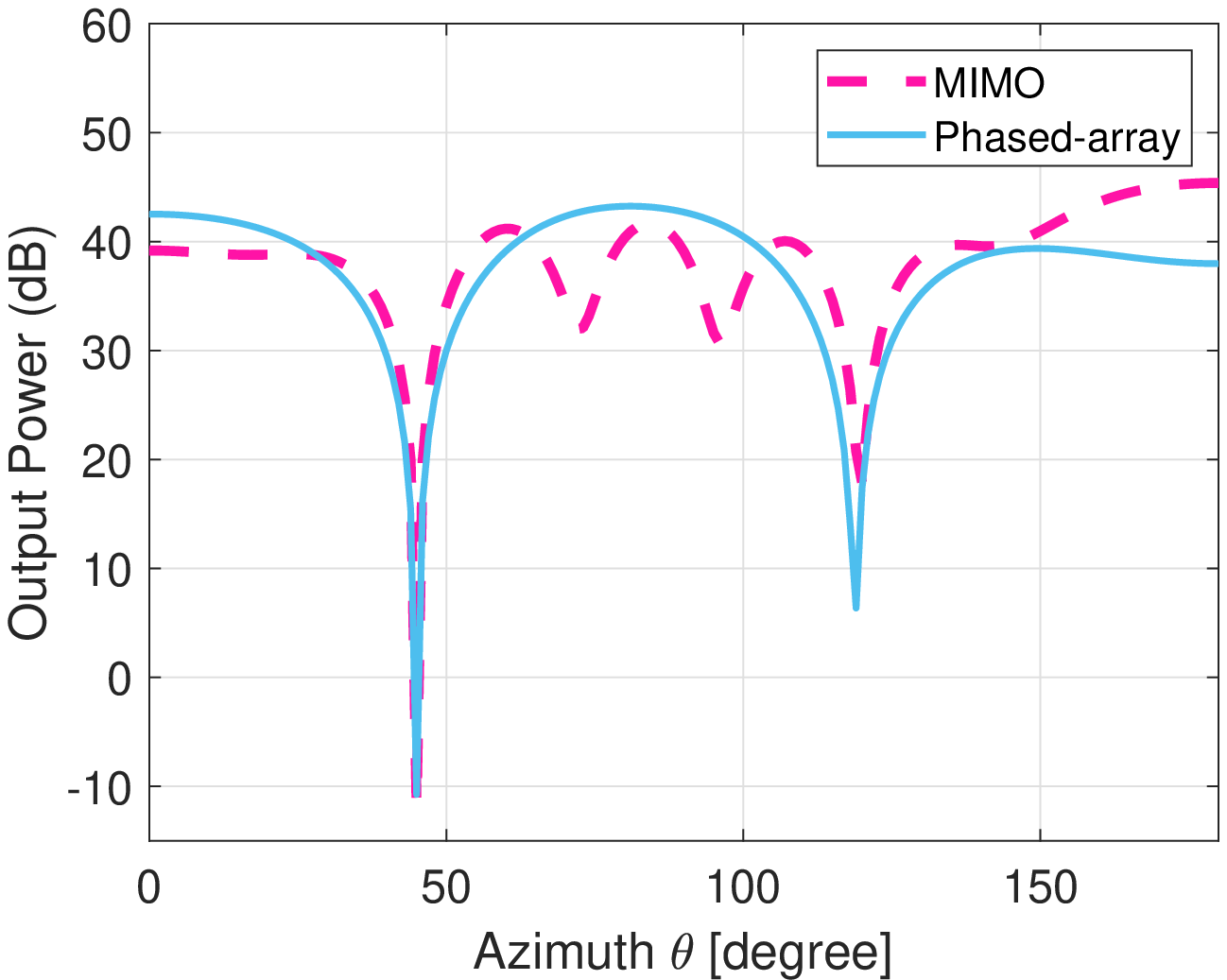}}
	\subfigure[]{
		\label{Fig.sub.4}
		\includegraphics[width=0.23\textwidth]{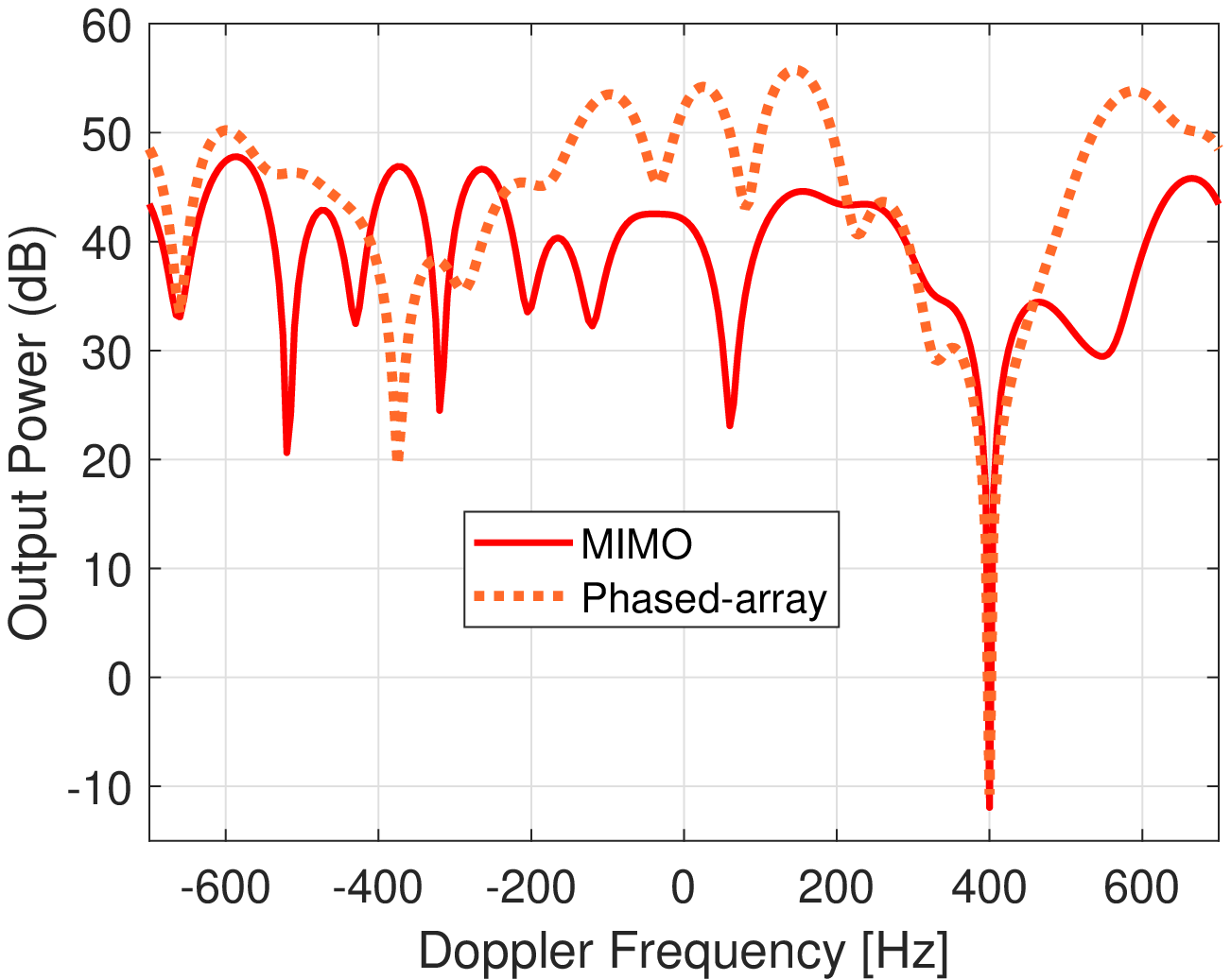}}
	\caption{The cuts of the MIMO and the phased-array adapted patterns (a) at the Doppler frequency ${{f}_{d}}=400 \kern 2pt Hz$, and (b) at the azimuth angle $\theta ={{45}^{o}}$.}
	\label{Fig.main222}
\end{figure}
\subsection{SINR Loss}
It is often useful to express the performance relative to what could be obtained in the absence of interferences. 
The SINR loss, ${{L}_{SINR}}\left( \psi ,{{f}_{d}} \right)$, is defined to be the processed performance relative to what could be obtained in an interference-free environment, i.e.,
\begin{equation}
{{L}_{SINR}}\left( \psi ,{{f}_{d}} \right)=20{{\log }_{10}}\left( \frac{\left| {{\mathbf{v}}^{H}}\mathbf{\bar{q}}\left( {{r}_{t}},\psi ,{{f}_{d}};\mathbf{w}={{\mathbf{1}}_{{{N}_{T}}\times 1}} \right) \right|}{{{N}_{T}}{{N}_{R}}L} \right).
\end{equation}
The gain of ${{N}_{T}}{{N}_{R}}L$ represents the coherent spatial and temporal integration over the virtual ${{N}_{T}}{{N}_{R}}$ elements and the $L$ pulses.
It may be noted that ${{L}_{SINR}}$ is a function of Doppler and azimuth. However, it is usually plotted as a function of Doppler, assuming that the array is oriented in the desired target direction.
Fig. 9 shows the SINR loss curves obtained by the FDA, MIMO, and phased-array systems, where the azimuth angle is fixed at $\theta ={{90}^{o}}$.
As expected, when the target velocity is close to zero Doppler, the performance loss of the FDA is very small as compared to the MIMO and the phased-array systems.
\begin{figure}
	\centering
	\includegraphics[width=0.45\textwidth]{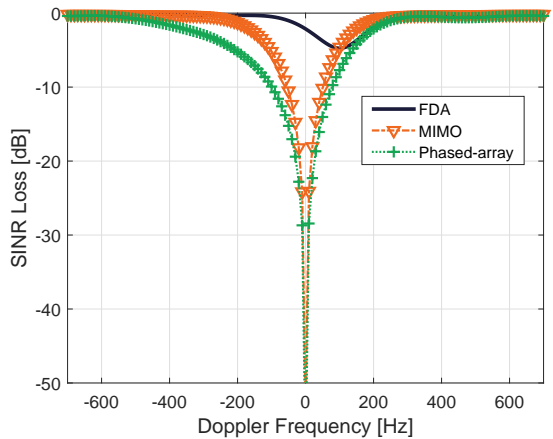}
	\caption{The SINR loss curves obtained by FDA, MIMO and phased-array systems.}
\end{figure}

\section{Conclusion}
In this paper, using range-controllable degrees of freedom, we present a method for high-gain detection of an airborne FDA radar for targets obscured by mainlobe clutter with the same Doppler frequency.
In order to do so, a receiver for coherent FDA is proposed that takes full advantage of the orthogonality of the echo signals in the Doppler domain.
For each receive antenna, a joint space-time processing module is applied to extract the desired signals from the aliased returns.
Then, the receive signal model considering the ground clutter and the barrage jamming is devised.
Finally, the target detection is formed using an adapted MVDR spatial-temporal receiver.
The proposed joint range-space-time processing method is also easily extendable with additional constraints that may occur in practice, although even so, the output SINR can be expected to improve by jointly designing the transmit and receive weights.

% \appendices
\appendix
Using the equality
\begin{equation}
\begin{aligned}
{{\mathbf{p}}_{M\times 1}}\otimes {{\mathbf{b}}_{N\times 1}}&={{\left( {{\left( {{\mathbf{b}}_{N\times 1}} \right)}^{H}}\left[ {{\left( {{\mathbf{p}}_{M\times 1}} \right)}^{H}}\otimes {{\mathbf{I}}_{N}} \right] \right)}^{H}} \\ 
& =\left( {{\mathbf{p}}_{M\times 1}}\otimes {{\mathbf{I}}_{N}} \right){{\mathbf{b}}_{N\times 1}} \\ 
\end{aligned}
\end{equation}
where ${{\mathbf{p}}_{M\times 1}}$ and ${{\mathbf{b}}_{N\times 1}}$ denote the $M\times 1$- and ${N\times 1}$-dimensional vectors, respectively,
the vector $\mathbf{\bar{q}}\left( r,\psi ,{{f}_{d}};\mathbf{w} \right)$ can be equivalently expressed as
\begin{equation}
\begin{aligned}
& \mathbf{\bar{q}}\left( r,\psi ,{{f}_{d}};\mathbf{w} \right) \\ 
& ={{\mathbf{b}}_{dop}}\left( {{f}_{d}} \right)\otimes \left( {{\mathbf{a}}_{R}}\left( \psi  \right)\otimes \left[ \mathbf{w}\odot {{\mathbf{a}}_{T}}\left( r,\psi  \right) \right] \right) \\ 
& =\left( {{\mathbf{b}}_{dop}}\left( {{f}_{d}} \right)\otimes {{\mathbf{I}}_{{{N}_{T}}{{N}_{R}}}} \right)\left( {{\mathbf{a}}_{R}}\left( \psi  \right)\otimes \left[ \mathbf{w}\odot {{\mathbf{a}}_{T}}\left( r,\psi  \right) \right] \right) \\ 
& =\left( {{\mathbf{b}}_{dop}}\left( {{f}_{d}} \right)\otimes {{\mathbf{I}}_{{{N}_{T}}{{N}_{R}}}} \right)\left( {{\mathbf{a}}_{R}}\left( \psi  \right)\otimes {{\mathbf{I}}_{{{N}_{T}}}} \right)\left[ \mathbf{w}\odot {{\mathbf{a}}_{T}}\left( r,\psi  \right) \right] \\ 
& =\mathbf{C}\left( r,\psi ,{{f}_{d}} \right)\mathbf{w} \\ 
\end{aligned}
\end{equation}
where 
\begin{equation}
\begin{aligned}
& \mathbf{C}\left( r,\psi ,{{f}_{d}} \right)=\left( {{\mathbf{b}}_{dop}}\left( {{f}_{d}} \right)\otimes {{\mathbf{I}}_{{{N}_{T}}{{N}_{R}}}} \right) \\ 
&\kern 20pt \times \left( {{\mathbf{a}}_{R}}\left( \psi  \right)\otimes {{\mathbf{I}}_{{{N}_{T}}}} \right)\operatorname{diag}\left\{ {{\mathbf{a}}_{T}}\left( r,\psi  \right) \right\}. \\ 
\end{aligned}
\end{equation}
Substituting $(66)$ into $(54)$ yields
\begin{equation}
\psi \left( \mathbf{v},\mathbf{w} \right)=\text{SNR}\cdot \frac{{{\mathbf{w}}^{H}}{{\mathbf{C}}_{t}}\left( \mathbf{v} \right)\mathbf{w}}{{{\mathbf{w}}^{H}}{{\mathbf{C}}_{c}}\left( \mathbf{v} \right)\mathbf{w}+\eta \left( \mathbf{v} \right)}
\end{equation}
where
\begin{subequations}
	\begin{equation}
\kern -58pt {{\mathbf{C}}_{t}}\left( \mathbf{v} \right)={{\mathbf{C}}^{H}}\left( {{r}_{t}},{{\psi }_{t}},{{f}_{t,d}} \right)\mathbf{v}{{\mathbf{v}}^{H}}\mathbf{C}\left( {{r}_{t}},{{\psi }_{t}},{{f}_{t,d}} \right)
	\end{equation}
	\begin{equation}
\begin{aligned}
& {{\mathbf{C}}_{c}}\left( \mathbf{v} \right)= \\ 
& \sum\limits_{i}^{I}{\sum\limits_{j}^{J}{\text{CN}{{\text{R}}_{i,j}}{{\mathbf{C}}^{H}}\left( {{r}_{c,j}},{{\psi }_{c,j}},{{f}_{c,d,j}} \right)\mathbf{v}{{\mathbf{v}}^{H}}\mathbf{C}\left( {{r}_{c,j}},{{\psi }_{c,j}},{{f}_{c,d,j}} \right)}} \\ 
\end{aligned}
	\end{equation}
\begin{equation}
\begin{aligned}
& \eta \left( \mathbf{v} \right)={{\mathbf{v}}^{H}}\left( {{\mathbf{I}}_{L}}\otimes \left[ \sum\limits_{{\tilde{j}}}^{{\tilde{J}}}{\text{JN}{{\text{R}}_{{\tilde{j}}}}{{\mathbf{a}}_{R}}\left( {{\theta }_{{\tilde{j}}}} \right)\mathbf{a}_{R}^{H}\left( {{\theta }_{{\tilde{j}}}} \right)} \right]\otimes {{\mathbf{I}}_{N_T}} \right)\mathbf{v} \\ 
& \kern 31pt +{{\mathbf{v}}^{H}}{{\mathbf{I}}_{N_TN_RL}}\mathbf{v} \\ 
\end{aligned}
\end{equation}
\end{subequations}
which completes the proof.

\bibliographystyle{IEEEtran}
\bibliography{Refs}

\end{document}